\documentclass[10pt,nofootinbib,twocolumn,aps,letterpaper,superscriptaddress,showpacs,prd,longbibliography]{revtex4-1}
\usepackage{amsmath}
\usepackage{amsfonts}
\usepackage{amssymb}
\usepackage{bbm}
\usepackage[dvips]{graphicx}
\usepackage[dvips,bookmarks=false]{hyperref} 

\hypersetup{pdfstartview=FitH,pdfhighlight=/O,colorlinks=false}

\begin{document}                

\newcommand{\mc}[1]{\mathcal{#1}}
\newcommand{\mbb}[1]{\mathbbm{#1}}
\newcommand{\Tr}{{\rm Tr}}

\title{\Large \bf Coherent states for FLRW space-times in loop quantum gravity}

\author{Elena Magliaro}
\email{magliaro@gravity.psu.edu}
\affiliation{Institute for Gravitation and the Cosmos, Physics Department, Penn State, University Park, PA 16802-6300, USA}
\author{Antonino Marcian\`o}
\email{amarcian@haverford.edu}
\affiliation{Department of Physics and Astronomy, Haverford College, Haverford, PA 19041, USA}
\author{Claudio Perini}
\email{perini@gravity.psu.edu}
\affiliation{Institute for Gravitation and the Cosmos, Physics Department, Penn State, University Park, PA 16802-6300, USA}

\date{\small\today}

\begin{abstract}


\noindent We construct a class of coherent spin-network states that capture properties of curved space-times of the Friedmann-Lama\^itre-Robertson-Walker type on which they are peaked. The data coded by a coherent state are associated to a cellular decomposition of a spatial ($t=$const.) section with dual graph given by the complete five-vertex graph, though the construction can be easily generalized to other graphs. The labels of coherent states are complex $SL(2, \mathbbm{C})$ variables, one for each link of the graph and are computed through a smearing process starting from a continuum extrinsic and intrinsic geometry of the canonical surface. The construction covers both Euclidean and Lorentzian signatures; in the Euclidean case and in the limit of flat space we reproduce the simplicial 4-simplex semiclassical states used in Spin Foams.
\end{abstract}


\maketitle

\section{Introduction}
Semiclassical states are a standard tool to select the semiclassical regime of a quantum theory. The semiclassical states in the Hilbert space of quantum General Relativity are states that are able to reproduce a given classical geometry in terms of their expectation values, and in which the quantum mechanical fluctuations are small. Within canonical Loop Quantum Gravity \cite{Ashtekar:1986yd,Rovelli:1989za,Ashtekar:1991kc,Rovelli:2004tv,Thiemann:2007zz} and Loop Quantum Cosmology \cite{Bojowald:2003md,Ashtekar:2003hd,Bojowald:2001xe,Bojowald:2002gz,Ashtekar:2007tv}, the use of semiclassical states has revealed fruitful in a number of applications, such as the analysis of the quantum constraints \cite{Thiemann:2000bx,Bahr:2006hm} and the computation of effective Hamiltonians \cite{Sahlmann:2002qj,Sahlmann:2002qk}. In the covariant Spin Foam setting  \cite{Reisenberger:2000fy,Perez:2003vx,Ashtekar:2010ve}, coherent states have been useful for understanding the correct way of implementing the constraints of BF-like theories \cite{Livine:2007vk,Engle:2007wy,Freidel:2007py}, while addressing their low-energy limit \cite{Magliaro:2007nc,Alesci:2008ec,Alesci:2008un,Conrady:2008mk,Barrett:2009gg,Barrett:2009mw} or investigating their renormalizability \cite{Perini:2008pd,Krajewski:2010yq,Geloun:2010vj}.

In the framework of the boundary formalism for generally covariant field theories \cite{Oeckl:2003vu}, a strategy to derive scattering amplitudes in Spin Foams  has been delined in \cite{Rovelli:2005yj,Modesto:2005sj}. The key idea is to use semiclassical states of geometry as a `background' for local measurements. For example, the semiclassical 2-point function can be computed, and the result has been compared to the standard graviton propagator on Minkowski space \cite{Bianchi:2006uf,Alesci:2007tx,Bianchi:2009ri}. Understanding the form of semiclassical states also for curved space-times is important for the generalization of n-point functions to curved backgrounds.

The calculation of semiclassical n-point functions are made asymptotically for large distance scales, to first order in a graph expansion, and to first order in the spin foam vertex expansion, so that only a finite set of degrees of freedom of Generaly Relativity is captured. A similar graph expansion has also been advocated in contexts of cosmological interest. This is the way in which a ``triangulated loop quantum cosmology'' has been derived \cite{Rovelli:2008dx,Battisti:2009kp,Marciano:2010jc,Battisti:2010he,Bianchi:2010zs} by means of such a graph truncation, directly from the full theory. The resulting expansion is neither an ultraviolet nor an infrared truncation, but it is rather equivalent to a mode expansion to the simplest modes of the gravitational field on a compact space. For example, in an almost homogeneous and isotropic universe, the lowest mode is represented by the scale factor $a(t)$. See \cite{Rovelli:2010wq} for a recent discussion on the rationale of this heuristic approximation.

In this paper, we present a class of coherent states useful for a semiclassical analysis on a spatially closed Friedmann-Lama\^itre-Robertson-Walker (FLRW) background. We follow the line of \cite{Bianchi:2009ky} (see also \cite{Dasgupta:2003da}) for the general construction and the relation between canonical and covariant semiclassical states, \cite{Battisti:2009kp} for the Maurer-Cartan formalism, and \cite{Bianchi:2010zs} for a similar application of coherent states to cosmology.

There is a simple way to construct a coherent state peaked on a given classical space-time, the logic is the following. Consider a space-like hypersurface $\Sigma_t$ of constant time in a closed FLRW space-time. $\Sigma_t$ has the topology of the 3-sphere. Take a regular cellular decomposition of $\Sigma_t$ and associate to it its dual graph. We will choose a regular geodesic graph with five nodes. This decomposition provides us with a set of curves and surfaces to be used for the smearing process. We first compute the holonomies $h_l$ of the Ashtekar connection along curves $l$ and fluxes $X_l$ of gravitational electric fields through the surfaces $S_l$ dual to $l$. The variables $h_l,X_l$ parametrize a truncation of the phase space of classical General Relativity. They can be used as semiclassical labels over which the coherent state is peaked. Equivalently, the polar decomposition
\begin{align}
H_l=h_l\,e^{X_l}\in SL(2,\mathbbm C)
\end{align}
constitutes the label of coherent states, one per each curve considered.

Those labels can be expressed, alternatively, in terms of a positive parameter $\eta$, an angle $\xi$, and two unit vectors $\vec n$,
\begin{align}
\{\eta_l, \xi_l,\vec n_{s(l)}, \vec n_{t(l)}\}\label{twisted labels}.
\end{align}
This geometrical parametrization of the phase space is the one of twisted geometries \cite{Freidel:2010aq,Rovelli:2010km,Freidel:2010bw}.

The parametrization \eqref{twisted labels} is used to compute the asymptotic expansion in the usual spin-network basis. Using the result \cite{Bianchi:2009ky}, this is given by a Gaussian distribution over spins $j$, times a phase factor that codes the extrinsic curvature of the slicing:
\begin{align}
e^{-\frac{(j-j_0)^2}{2\sigma^2}}\times e^{- i \xi j}.
\end{align}

In the next section we review the heat-kernel technique for coherent states in Loop Quantum Gravity. In section \ref{sec:CSN} we outline the main properties of FLRW geometry which are relevant to our construction. In section \ref{sec:CST} we compute the non-local observables associated to a given cellular decomposition. Those are the labels of the FLRW coherent states, discussed in section \ref{sec:QS}, where we determine their large scale behavior. We set the speed of light $c=1$ throughout this paper.
\section{Coherent spin-networks}\label{sec:CSN}
In LQG the kinematical Hilbert space associated to a graph  $\Gamma$, embedded in a spatial hyper-surface $\Sigma$, is $\mc H_\Gamma=L^2(SU(2)^L/SU(2)^N)$, where $L$ is the number of links of the graph and $N$ the number of its nodes. Kinematical states are then functions of $SU(2)$ group elements $h_l$ that are invariant under $SU(2)$ transformations at nodes, \begin{equation}
\Psi(h_l)=\Psi(g_{s(l)}h_lg_{t(l)}^{-1}),
\end{equation}
where $s(l)$ and $t(l)$ are respectively the nodes that are source/target of the link $l$. The standard orthonormal basis is labeled by spins $j_l$ associated to links and invariant tensors $i_n$ (intertwiners) associated to nodes; it is formed by spin-network states
\begin{align}
\Psi_{j_l,i_n}(h_l)=\otimes v_{i_n}\cdot\otimes D^{j_l}(h_l)
\end{align}
where $i_n$ labels an orthonormal set of intertwiners, $D^{j_l}$ are spin-$j_l$ unitary representation matrices and $\cdot$ denotes indices contraction.

Once a graph $\Gamma$ is fixed, spin-network states capture a finite number of d.o.f. of General Relativity: the ones associated to the classical phase space of holonomies of the Ashtekar-Barbero connection along links of the graph {\bf $\Gamma$} and fluxes through surfaces dual to the links of the graph {\bf $\Gamma$}. Now choose a classical configuration of the Ashtekar connection $A$ and its conjugate momentum, the gravitational electric field $E$. Moreover, let $\Delta_\Sigma$ be a cellular decomposition of $\Sigma$ and $\Gamma$ the graph which is the 1-skeleton of a dual complex $\Delta^*_\Sigma$. This provides a discretization of the manifold; fields are discretized smearing $A$, which is a $su(2)$-valued connection 1-form, and $E$, which is a $su(2)$-valued density vector, over curves and surfaces (the links of $\Gamma$ and the dual surfaces).

The connection is smeared along half-link $l$ of the graph $\Gamma$, that is from the source node $s(l)$ to the point of intersection with the surface. So we denote with $h_l$ the path-ordered exponential 
\begin{align}
h_l=P\exp\int_l A
\end{align}
which, implicitly, will be \emph{always} defined on half of the link $l$.
For this analysis we consider the following definition of the flux \cite{Thiemann:2000bv}: 
\begin{equation}
E_l=E(S_l)=\int_{S_l} Ad_U *E.
 \label{fluxdef}  
\end{equation}
Here the densitized inverse triad $E$ is parallel-transported by the holonomy $U$ to the integration point. $Ad$ stands for the action of the adjoint representation of $SU(2)$ on Lie algebra elements. $*$ is the Hodge dual operator. Definition \eqref{fluxdef} depends on the point $\sigma^0\in S_l$ that is used as base-point for the holonomies $U$.  This is chosen as the intersection-point between the link $l$ and the dual surface $S_l$.  The holonomy $U$ is computed along a path which starts at $\sigma_0$ and ends at the integration point $\sigma$. 
 The reason for considering this definition of the flux variable is the simple behavior under local $SU(2)$ gauge transformations:
  \begin{equation} 
  E(S)\rightarrow Ad_{G(\sigma^0)} E(S).
 \end{equation}

The set of couples $(h_l,E_l)$, one per each link of the graph, can be viewed as a point in a truncation of the phase space of General Relativity as captured by the graph $\Gamma$. The smeared Poisson algebra reads
\begin{align}
&\{U_l,U_{l'}\}=0,\,\, \quad \{E^i_l,E^j_{l'}\}=\delta_{ll'} \epsilon^{ijk} E^k_l,\nonumber\\\label{poissonsmeared}
&\{E^i_{l},U_{l'}\}=\pm\delta_{ll'}\,8\pi G\hbar\gamma\;\tau^iU_l,
\end{align}
derived from the fundamental brackets 
\begin{align}
&\{A_a^i(x),A_b^j(y)\}=0,\,\, \quad \{E^a_i(x),E^b_j(y)\}=0, \nonumber\\
&\{A_a^i(x),E^b_j(y)\}=8\pi G\gamma \,\delta^i_j\delta_a^b\delta(x,y),
\end{align}
where the non-vanishing real number $\gamma$ is the Barbero-Immirzi parameter. In the previous equations, $\tau^i=i\sigma^i/2$ are $su(2)$ generators defined in terms of the Pauli matrices $\sigma^i$. The sign $\pm$ in \eqref{poissonsmeared} depends on the relative orientation between the link $l$ and the surface $S_l$. In the following we will choose the `$+$' orientation. The couple $(h_l,E_l)$ can be identified with an element of  $SL(2,\mbb{C})$, the complexification of $SU(2)$, using the polar decomposition 
\begin{equation}
H_l=\;h_l\;\exp(i\frac{\alpha_lE_l}{8\pi G \hbar \gamma}).
\label{eq:complexification}
\end{equation}
Coherent spin-networks with labels as in (\ref{eq:complexification}) are peaked on the classical configuration $(h_l,E_l)$. The presence of the positive real numbers, called heat-kernel times, $\alpha_l$ in (\ref{eq:complexification}) will become clearer later on (see equation \eqref{j0final} and the comment following it).

The construction of coherent states for quantum gravity relies on a heat-kernel technique, that in the following lines we first review for the simple example of a quantum particle in non-relativistic mechanics. Consider the heat-kernel of $L^2(\mathbbm R^n,d x)$ defined by:
\begin{align}
K_t(x,x')=e^{-\frac{\alpha}{2}\Delta_x}\delta(x,x')
\end{align}
where $\Delta_x$ is the Laplacian on $\mathbbm R^n$. The phase space of a particle in $\mathbbm R^n$ is $\mathbbm R^{2n}\simeq\mathbbm C^n$, the complexification of the Abelian group $\mathbbm R^n$. 
Consider now the unique analytic continuation of the heat-kernel w.r.t. the variable $x'$. We have thus defined the family of wave functions
\begin{align}
\psi^\alpha_{z}(x)=K_\alpha(x,z)\quad\quad z\in\mathbbm C^n.
\end{align}
Those states are \emph{coherent} in the following mathematical sense:
\begin{itemize}
\item They are eigenstates of the annihilation operator $$\hat z=\hat x+i\alpha\hat p,$$
\item saturate the Heisenberg uncertainty relation $$\Delta x\Delta p=\frac{\hbar}{2},$$
\item form an overcomplete basis of $L^2(\mathbbm R^n,d x)$
\begin{align}\nonumber
\int\overline{\psi^\alpha_z(x)}\psi^\alpha_{z}(x')d z=\delta(x,x').
\end{align}
\end{itemize}

We are interested in coherent states for the sector of LQG associated to a single graph $\Gamma$. The main ingredient are Hall coherent states \cite{Hall1,Hall2}, generalization of the previous construction from the abelian group $\mathbbm R^n$ to a general compact Lie group. We restrict our attention to $SU(2)$. First, apply the heat-kernel evolution to the Dirac delta distribution over the group:
\begin{align}
K_\alpha(h,h')=e^{-\frac{\alpha}{2}\Delta_h}\delta(h,h')
\label{Deltatodelta}
\end{align}
where the Laplace-Beltrami operator $\Delta_h$ on $SU(2)$ is defined w.r.t. the unique bi-invariant metric tensor. Explicitely, we have
\begin{equation}
K_\alpha(h,h')=\sum_j (2j+1) e^{-j(j+1)\frac{\alpha}{2}}\; \Tr D^{(j)}(h^{-1}h').
\end{equation}
Now take the unique analytic continuation of \eqref{Deltatodelta} w.r.t. the variable $h'$, which defines wave-functions $\psi^\alpha_H(h)$ labeled by an element $H$ in the complexification $SU(2)^\mathbbm C$, which is $SL(2,\mathbbm C)$:
\begin{align}
\psi^\alpha_{H}(h)=K_\alpha(h,H)\quad\quad H\in SL(2,\mathbbm C).
\end{align}
Being $SU(2)$ simply connected, $SU(2)^\mathbbm C$ is defined via exponentiation of the complexification of the Lie algebra.
Intuitively, the heat-kernel technique is the natural way to construct `Gaussian' wave-packets on $SU(2)$.

Applying the heat-kernel technique to several copies of $SU(2)$ allows to build coherent spin-network states for LQG \cite{Ashtekar:1994nx,Thiemann:2000bw,Thiemann:2000ca,Bahr:2007xn,Bianchi:2009ky}.
Coherent spin-networks are defined as follows: we consider the gauge-invariant projection of a product over the links of a graph of heat kernels,
\begin{equation}
\Psi_{\Gamma,H_l}(h_l)=\int\prod_n dg_n\,\prod_{l} K_{\alpha_{l}}(h_{l},\,g_{s(l)}\, H_{l}\, g_{t(l)}^{-1}),
\label{coherent SN}
\end{equation}
where we have a $SU(2)$ integration for each node $n$. 
Here, $\alpha_{l}$ are positive real numbers (heat-kernel times) that can be fixed from some dynamical requirement. As shown in \cite{Ashtekar:1994nx}, coherent spin-networks provide a Segal-Bargmann transform for Loop Quantum Gravity, that has been lifted to Spin Foams in \cite{Bianchi:2010mw}.

We can use a parametrization of  $SL(2,\mathbbm C)$ with an interpretation in terms of discrete geometries. Any $SL(2,\mathbbm C)$ element $H_{l}$ can be written as
\begin{align}
H_{l}=g_{\vec n_{s(l)}}  e^{(\eta_l+i\xi_l)\frac{\sigma_3}{2}} g^{-1}_{\vec n_{t(l)}}
\label{fromHtoFS}
\end{align}
that is as a $SU(2)$ rotation that brings the direction $\vec n_{t(l)}$ on the direction $\vec z=(0,0,1)$ times a $SL(2,\mathbbm C)$ transformation along $\vec z$ times a rotation that brings $\vec z$ on $\vec n_{s(l)}$. This decomposition is unique once we choose a map $S^2\rightarrow SU(2)$ at each node, namely a section of the Hopf fibration. 

A different choice for those sections implies a redefinition (shift) of the parameters $\xi_l$. Notice that, while this choice is purely conventional, a shift of $\xi_l$ that keeps \emph{fixed} the section will change $H_l$. But the physical information is contained in $H_l$. We will see how $H_l$, hence $\xi_l$, is determined unambiguously from the FLRW geometry.

The decomposition \eqref{fromHtoFS}, discussed in \cite{Bianchi:2009ky}, provides the following equivalent set of labels for coherent states:
\begin{align}
\{\eta_l, \xi_l,\vec n_{s(l)}, \vec n_{t(l)}\}\label{twisted labels2}
\end{align}
i.e. a positive real number, an angle and two unit vectors. The parameter $\eta_l$ is related to the modulus of the gravitational flux through the surface which is intersected by the link, namely to the area of a surface. The unit vector $\vec n_{s(l)}$ is interpreted as the unit-flux, parallel transported at the source node (and similarly for $\vec n_{t(n)}$). The angle $\xi_l$ is the conjugacy class over which the holonomy of the Ashtekar-Barbero connection is peaked, and therefore it codes the extrinsic curvature.

In terms of those variables, the following large distance (large $\eta$) asymptotic behavior for coherent spin-networks can be found\cite{Bianchi:2009ky}\footnote{We are omitting an overall normalization factor.}:
\begin{align}
\Psi_{H_l}(h_l)\simeq\sum_{j_l,i_n} \prod_l e^{-\frac{(j_l-j^0_l)^2}{2 \sigma^2_l}}e^{-i \xi_l j_{l}}(\prod_n \Phi_{i_n})\Psi_{j_{l},i_n}(h_l).
\label{asymptotics}
\end{align}
This is a Gaussian with phase. The position of the peak $j_l^0$ is related to $\eta_l$ by $(2 j_l^0+1)=2 \eta_{l}/\alpha_{l}$, and the spread of the Gaussian around $j_l^0$ is governed by the parameter $\sigma_{l}=1/\sqrt{\alpha_{l}}$. Finally, $\Phi_{i_n}$ is the coefficient for the expansion of the Livine-Speziale coherent intertwiner \cite{Livine:2007vk} on a orthonormal basis labeled  by $i_n$, and carries the dependence on the unit vectors.
\section{FLRW: classical space-time}\label{sec:CST}

In this section we review some properties of FLRW space-time, that will be useful for the application to coherent states in Loop Quantum Gravity. We consider a time-oriented globally hyperbolic space-time with topology $\mathbbm R\times S^3$ and line element
\begin{equation}
ds^2=-dt^2+a(t)^2d\Omega
\end{equation}
where the function $a(t)$ is the scale factor and
\begin{align}
d\Omega=d\psi^2+\sin^2\psi(d\theta^2+\sin^2\theta d\phi^2)
\end{align}
is the metric of the Euclidean 3-sphere. Here $\theta\in [0,\pi),\phi\in[0,2\pi),\psi\in[0,2\pi)$, namely, we are using hyperspheric coordinates. This geometry describes a homogeneous and isotropic expanding or contracting universe. We want to construct a semiclassical state as in \eqref{coherent SN} which is peaked on the intrinsic \emph{and} the extrinsic geometry  
of a spatial ($t=$const.) section of FLRW space-time.

For calculation purposes, we shall use the Maurer-Cartan formalism for homogeneous spaces, as done in \cite{Battisti:2009kp}. The unit Euclidean 3-sphere is diffeomorphic to the Lie group $SU(2)$. The manifold $SU(2)$ is then an homogeneous space w.r.t. its own action, the latter being free and transitive. It carries a natural homogeneous (left-invariant) $su(2)$-valued form, named Maurer-Cartan form,
\begin{align}
\omega=g^{-1}dg=\omega^i_a \tau^i dx^a
\end{align}
which satisfies the structural equation 
\begin{align}
d\omega^i+\frac{1}{2}\,\epsilon^{i}_{\,\,jk}\,\omega^j\wedge\omega^k=0,
\end{align}
namely $\omega_a^i$ also defines a flat principal connection over $SU(2)$.
The spatial sections are described by a time-dependent 3-dimensional metric tensor that can be written in terms of the Maurer-Cartan form, the latter viewed as a frame field (a cotriad): 
\begin{align}
g_{ab}(t)=a(t)^2 \omega^i_a \omega^i_b
\end{align}
More precisely, the cotriad for a universe of radius $a(t)$ is 
\begin{align}
e_a^i=a(t) \omega^i_a.
\end{align}
This corresponds to a specific class of gauge fixing which makes the cotriad proportional to the Maurer-Cartan 1-form.
The triad is the dual vector field
\begin{align}
e^{ai}=g^{ab}e^i_b.
\end{align}

The explicit expression in hyperspheric coordinates can be found in the Appendix.
The Ashtekar-Barbero connection $A=A_a^i \tau^i dx^a$ has components
\begin{align}
A_a^i=\Gamma_a^i+\gamma K_a^i
\end{align}
with $\Gamma_a^i$ the spin-connection and $K_a^i$ the extrinsic curvature. It can be written in a homogeneous gauge where it is left-invariant and proportional to the Maurer-Cartan connection
\begin{align}
A_a^i=(\Gamma+\gamma K)\omega_a^i.
\label{scalarcoefficients}
\end{align}
To compute the scalar coefficients $\Gamma$ and $K$, we first write $\Gamma_a^i=\Gamma \omega_a^i$; the proportionality coefficient $\Gamma$ can be computed by first evaluating the Ricci scalar, and comparing with the known value for the 3-sphere of radius $a(t)$, namely
\begin{align}
R=\frac{6}{a(t)^2}=\epsilon^{ijk}e_a^j e_b^k (D\Gamma)^i_{ab},
\end{align}
where $D$ is the covariant exterior derivative. This fixes the intrinsic curvature coefficient in \eqref{scalarcoefficients} as $\Gamma=1/2$. The extrinsic curvature is (half) the Lie derivative with respect to the unit future-oriented vector field $\partial/\partial t$ normal to the space-like surface,
\begin{align}
K_{ab}=\frac{1}{2}\mathcal L_{\frac{\partial}{\partial t}} g_{ab}=\frac{1}{2}\frac{\partial}{\partial t}g_{ab}=a\dot a \,\omega_a^i\omega_b^i
\end{align}
so that, raising one index by means of the inverse triad field, we get
\begin{align}
K_a^i=e^{bi}K_{ab}=\dot a\,\omega_a^i.
\end{align}
The last relation fixes the scalar coefficient $K$ of the extrinsic curvature in \eqref{scalarcoefficients} to be $K=\dot a(t)$. 
\section{FLRW: cellular decomposition}\label{sec:CD}
Consider a cellular decomposition $\Delta$ of the constant time 3-surface $\Sigma_t$, defined as follows. In the  Euclidean 3-sphere of radius $a(t)$ take five equally spaced points, and join them with ten geodesic paths. We obtain an embedded complete graph with 5 vertices, 1-skeleton of $\Delta$. Every closed loop joining three points is the boundary of a minimal surface, which is a totally geodesic triangle, and a 2-cell of $\Delta$. The 3-cells are the closed regions bounded by four mutually adjacent 2-cells. We need also the dual complex $\Delta^*$, isomorphic to $\Delta$, whose vertices are barycenter of the 3-cells of $\Delta$. Call $\Gamma_5$ the 1-skeleton graph of $\Delta^*$ where $l$ labels its links. Each link $l$ of $\Gamma_5$ cuts exactly one surface $S_l$ of $\Delta$ through the barycenter.

\begin{figure}
\includegraphics[width=3cm]{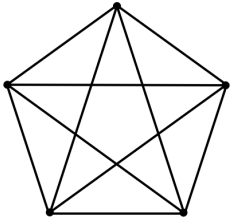}
\caption{The complete graph with 5 nodes, 1-skeleton of the cellular decomposition $\Delta\simeq\Delta^*$.}
\label{Fig:4-simplex}
\end{figure}

The cellular decomposition and its dual, in particular the surfaces $S_l$ and the dual links $l$, constitute the structure needed for the smearing process. 
\subsection*{Computation of holonomies}
We have to compute holonomies of the left-invariant Ashtekar-Barbero connection $A_a^i=c\,\omega_a^i$, with 
\begin{align}
c=\Gamma+\gamma K.
\end{align}
This task is easily accomplished if the path is a geodesic, as in our case.

Recall that the holonomy of the connection $A$ along the curve $\gamma$ is the path-ordered exponential\footnote{The holonomy of the $su(2)$ connection $A$ associated to a parametrized curve $\gamma(s)$, $s\in[0,s_0]$, is the solution evaluated at $s=s_0$ of the $SU(2)$ matrix differential equation
\begin{equation}
\begin{cases}\frac{d}{ds} U(s) + \dot{\gamma}^a (s)A_a(\gamma(s)) U(s)= 0\vspace{0.15cm}\\
U(0)=\mathbbm 1
\end{cases}
\end{equation}
}
\begin{align}
U(A)=\mathcal P\exp\int_\gamma A=\sum_{m=0}^\infty I_m,
\label{path-ordered}
\end{align}
where the $m$-th integral has the form
\begin{align}
I_m=\int_0^{L}\!\!\!\!d s_1\!\!\int_0^{s_1}\!\!\!\!\!\!ds_2\ldots\!\!\int_0^{s_{m-1}}\!\!\!\!\!\!\!\!\!\!\!\!\!ds_m \dot\gamma(s_1)\ldots\dot\gamma(s_m)A(s_1)\ldots A(s_m).
\end{align}
Here we have used an explicit parametrization of the geodesic $\gamma(s)$ in terms of the proper distance $s$ along the curve ($g_{ab}\dot\gamma^a\dot\gamma^b=1$) and $L$ is the proper length of the curve. Now we exploit the fact that since $e^{ai},\,i=1\ldots 3$ are three left-invariant vector fields, and the spatial metric tensor $g_{ab}$ is right-invariant\footnote{Of course, the spatial metric $g_{ab}$ is both left and right-invariant. It is the unique bi-invariant metric tensor on $SU(2)\simeq S^3$, up to a global scale factor, which is fixed to be $a(t)$.}, $e^{ai}$ are Killing vectors \cite{Wald:1984rg}. It follows that the three scalars
\begin{align}\label{conserved}
e_a^i \dot\gamma^a\equiv n^i
\end{align}
are conserved quantities, i.e. constant along (spatial) geodesics. We can then easily compute the path-ordered exponential \eqref{path-ordered}. Given
\begin{align}
I_m=\frac{1}{m!} (\frac{L}{a(t)}\,\vec n\cdot\vec\tau)^m,
\end{align}
we have\footnote{Geodesics over $SU(2)\simeq S^3$ that start at the identity element have the simple form $g(s)=e^{s\vec n \cdot \vec \tau}$, namely they are the 1-parameter subgroups of $SU(2)$.}
\begin{align}
U_{\gamma}(c\,\omega)=U_\gamma(\omega)^c=e^{c\frac{L}{a(t)}\vec n\cdot\vec \tau}.
\label{holonomycomega}
\end{align}
Notice that the first equality in \eqref{holonomycomega} does not hold for any path connecting the initial and final points, but only for geodesic paths. In fact if it did work for general paths, since $U_{\gamma}(\omega)$ is path-independent, that would imply that the holonomy of the Ashtekar connection is path-independent, or equivalently, that the holonomy of any contractible loop is the identity (which means that the connection is flat). Instead, the Maurer-Cartan holonomy is given by formula \eqref{holonomycomega} with $c=1$, for any path.

Specifically, we are interested in holonomies along the oriented links of the embedded graph $\Gamma_5$. A link goes from the source node $s(l)$ to the target node $t(l)$ of the geodesic link $l$. In fact, as explained in Section \ref{sec:CSN}, we consider holonomies along half-links, from the source node to the point of intersection with the dual surface. 

Take a node of $\Gamma_5$ and suppose all the four surrounding links are oriented as `outgoing'. It is clear that since the four links emanate from the node in isotropic directions, the four unit vectors defined in \eqref{conserved} are such that $\vec n_{l}\cdot \vec n_{l'}=\arccos(-1/3)$ for $l\neq l'$. These can be thought as the unit vectors normal to the four faces of an equilateral tetrahedron in $\mathbbm R^3$. For the general case, observe that unit vectors associated to different orientations of the path are related by $\vec n_{l}=-\vec n_{l^{-1}}$. This fixes uniquely the full set of 10 unit vectors, up to a global rotation. 

Moreover the length of a full link is
\begin{align}
L=a(t)\,2\Theta\,,\quad\quad\Theta=\arccos(-1/4).
\label{full length}
\end{align}
This can be easily seen considering the following geodesic on the unit Euclidean sphere $S^3\simeq SU(2)$: 
\begin{align}\label{geodesicsimple}
\gamma(s)=e^{s\tau_3}=\left(\begin{array}{cc}e^{i s/2}&0\\0&e^{-i s/2}\end{array}\right).
\end{align}
Suppose we have chosen coordinates where this geodesic lies over one link $l$ of the graph. With the standard embedding\footnote{The standard embedding of $SU(2)$ in $\mathbbm R^4$ is
\begin{align}
\left(\begin{array}{cc}x_1+ix_2&x_3+ix_4\\-x_3+ix_4&x_1-ix_2\end{array}\right)\in SU(2),
\end{align}
with $x_1,\ldots,x_4$ real and $x_1^2+x_2^2+x_3^2+x_4^2=1$.} of $SU(2)$ in $\mathbbm R^4$, it becomes clear that the two nodes $N_{s(l)}$, $N_{t(l)}$, viewed as vectors in $\mathbbm R^4$, have scalar product $N_{s(l)}\cdot N_{t(l)}=\cos\Theta$. Now, since the previous geodesic \eqref{geodesicsimple}, embedded in $\mathbbm R^4$, has the form $N(s)=(\cos\frac{s}{2},\sin\frac{s}{2},0,0)$, imposing
\begin{align}
N(0)\cdot N(s)=\cos\frac{s}{2}=\cos\Theta,
\end{align}
we find the value $s=2\Theta$ for the geodesic length. To obtain the geodesic length for a sphere of different radius, we just multiply by the appropriate scale factor $a(t)$, so we prove \eqref{full length}.

Thus, we find that the holonomy of the Ashtekar-Barbero connection along half-link $l$ is
\begin{align}
U_{l}(A)=e^{(\Gamma+\gamma K)\Theta\,\vec n_{l}\cdot\vec\tau}.
\label{Ul}
\end{align}
This completes the computation of holonomies. Notice that the dependence on $t$ in the holonomy is contained in the extrinsic curvature coefficient $K$, that codes the embedding properties of the 3-sphere into the curved space-time.
\subsection*{Computation of fluxes}
The computation of fluxes is more tricky as it relies on the definition \eqref{fluxdef}. The flux $E(S)=E(S)^i\tau^i$ depends on the surface as well as on the holonomies along a system of paths, as explained in section \ref{sec:CSN}. However, we shall not need those complicated details, as we are mostly interested in the unit-fluxes, and not in the explicit calculation of the modulus. In spite of this fact, the smearing process must not break the symmetries of the regular cellular decomposition we have chosen.
In our case we can take a family of geodesics joining the intersection point $\sigma_0$ with the generic point $\sigma$ of integration on the surface. We have
\begin{align}
E^i(S)=
&\int_S n^i\det h \,\, d^2\sigma
\end{align}
where $h_{ab}$, $a,b=1,2$, is the metric induced on the surface from $g_{ab}$, and $n^i$ is a unit vector given by
\begin{align}
&n^i=\frac{N^i}{\sqrt{N^jN^j}},
\end{align}
with
\begin{align}
N^i(\sigma)=R_{\sigma_0\rightarrow \sigma}^{ij}\, e^{aj}(\sigma_0)n_a(\sigma).
\end{align}
Let us explain our short notation. The rotation matrix $R^{ij}_{\sigma^0 \rightarrow{} \sigma}$ is the holonomy in the adjoint representation of $SU(2)$, that acts on the internal indices. It performs the parallel transport of the triad from the base point $
\sigma_0$ to the point of integration $\sigma$, along a geodesic path. Since we are averaging $n^i$ around the barycenter $\sigma_0$, we have clearly
\begin{align}
E^i(S)=|E(S)|\,n^i(\sigma_0),
\end{align}
where $|E(S)|=\sqrt{E(S)^i E(S)^i}$ denotes the modulus of the flux, whose time dependence is easily recovered: $|E(S)|\propto a(t)^2$.
We denote $E_{l}=E(S_l)$ the flux across the oriented surface $S_l$ punctured by the link $l$. We have
\begin{align}
E_l=|E| \,\vec n_l\cdot\vec \tau.
\end{align}

It is important to remark that the unit fluxes $\vec n_l= \vec E_{l}/|E|$ in the last equation coincide with the unit directions that identify the 1-parameter subgroup of Ashtekar holonomies $U_l$, namely with  $\vec n_l$ of equation \eqref{Ul}. This is true because the orientations of the link $l$ and of the surface $S_l$ agree.
\section{FLRW: quantum state}\label{sec:QS}
We can now define the coherent spin-network for FLRW geometry as the one labeled by $SL(2,\mathbbm C)$ variables on links, as defined by the smearing process of previous section:
\begin{align}
H_l=U_l\,e^{X_l}\quad\quad X_l=\alpha_l E_l/\gamma.
\label{labelspolar}
\end{align}
We now apply the decomposition \eqref{fromHtoFS} to obtain:
\begin{align}
H_l=g_{\vec n_l}e^{(\Gamma+\gamma K)\Theta\tau_3+i|X| \tau_3}g_{\vec n_l}^{-1},
\label{decompositionfinal}
\end{align}
where $g_{\vec n_l}^{-1}$ is \emph{precisely} the inverse of $g_{\vec n_l}$, namely there is no extra relative phase. As we anticipated in section \ref{sec:CSN}, the smearing process determines unambiguously the relative phase between `source' and `target' $SU(2)$ holonomies. In the asymptotic regime, this translates into a precise prescription for the relative phases of Livine-Speziale intertwiners, as we shall see in a moment.

The term proportional to $\Gamma$ in the exponent of \eqref{decompositionfinal} can be absorbed in the redefinition of the arbitrary phase of one of the $g_{\vec n}$, that is
\begin{align}
H_l=g'_{\vec n_l}e^{(\gamma K)\Theta\tau_3+i|X| \tau_3}g_{\vec n_l}^{-1},
\label{labelsFS}
\end{align}
in which
\begin{align}
g'_{\vec n_l} =U_l(\Gamma)g_{\vec n_l}=g_{\vec n_l} e^{\Gamma\Theta\, \tau_3}.
\end{align}
Notice that this choice of the relative phase between the `source' and `target' $SU(2)$ group elements is analogous to (actually, in the asymptotic regime coincide with) the canonical choice of relative phase for Livine-Speziale coherent intertwiners in the boundary of a flat 4-simplex of references \cite{Barrett:2009gg,Barrett:2009mw}. There, the canonical relative phase is obtained by the parallel transport of coherent states using the \emph{discrete} spin-connection. Here instead the parallel transport $U_l(\Gamma)$ is computed as the holonomy of the \emph{smooth} spin-connection $\Gamma_a^i$ along geodesics of the 3-sphere.

The coherent spin-network with labels as in \eqref{labelspolar}, or equivalently \eqref{labelsFS}, can be written as a superposition over the ordinary spin-network orthonormal basis $\Psi_{j_l,i_n}$ as
\begin{align}
\Psi_{H_l}(h_l)=\sum_{j_l,i_n}\psi_{H_l}(j_l,i_n)\Psi_{j_l,i_n}(h_l).
\end{align}
By the asymptotic formula \eqref{asymptotics}, the asymptotic behavior of the coherent spin-network for large $|X|\propto a(t)^2$ can be found. We find that for large scale factor $a(t)$, the FLRW coherent state is
\begin{align}
\psi_{H_l}(j_l,i_n)\simeq \prod_{l}e^{-\frac{(j_l-j_0)^2}{2\sigma^2}}\,e^{i\gamma K\Theta j_l}\,\prod_n\Phi_{i_n}.
\label{asymptotic state}
\end{align}
with
\begin{align}
&j_0=\frac{|E|}{8\pi G\hbar\gamma}\label{j0final}\\
&K=\dot a(t)\\
&\Theta=\arccos(-1/4)
\end{align}
and we have set the inverse heat-kernel times $\sigma_l=\sigma$, to respect the symmetry of the regular cellular decomposition. Lastly, let us comment on the units in \eqref{j0final}: by the definition \eqref{eq:complexification}, the dependence on the heat-kernel time in $j_0$ drops out, and we are left with the dimensionful factor $8\pi G\hbar\gamma$. This is in agreement with the area spectrum of Loop Quantum Gravity. 
In the following we give specific examples.

Our construction applies to the case of flat space-time, provided that we consider the Riemannian signature $(++++)$ and space-time topology $\mathbbm R^4$. Indeed, in this case we are allowed to write the metric in polar coordinates in the form
\begin{align}
ds^2=dr^2+r^2 d\Omega_3, 
\label{stp}
\end{align}
where $d\Omega_3$, as usual, is the metric tensor of a unit Euclidean 3-sphere. Thus (\ref{stp}) has the FLRW form, provided by the scale factor
\begin{align}
a(r)=r,
\end{align}
and consistently with this latter relation the extrinsic curvature coefficient $K$, defined by $K^i_a=K\, \omega^i_a$, is given by
\begin{align}
K=\dot a(r)=1.
\end{align}
The  semiclassical state for Euclidean space-time is then characterized by the large scale behavior:
\begin{align}
\psi_{H_l}(j_l,i_n)\simeq \prod_{l}e^{-\frac{(j_l-j_0)^2}{2\sigma^2}}\,e^{i\gamma\arccos(-\frac{1}{4}) j_l}\prod_n\Phi_{i_n}.
\end{align}
Remarkably, those coefficients are similar\footnote{A significant difference is that the heat-kernel coherent states discussed here present (asymptotically) a diagonal spin-spin correlation matrix, while a non-diagonal correlation matrix seems to be required from matching conditions in the graviton propagator calculation \cite{Bianchi:2009ri}.} to those ones used in order to define correlation functions over flat space in the Spin Foam setting \cite{Rovelli:2005yj,Modesto:2005sj,Bianchi:2006uf,Alesci:2007tx,Bianchi:2009ri}. In particular, the oscillatory factor which prescribes the extrinsic curvature matches exactly with the analogous phase factor originally advocated by Rovelli's ansatz \cite{Rovelli:2005yj}. More precisely, it matches with the one of reference \cite{Bianchi:2009ri}, which includes the correct dependence on the Immirzi parameter. Moreover, in the Spin Foam setting, the angle $\Theta=\arccos(-1/4)$ is interpreted as a $4$-dimensional dihedral angle between two tetrahedra lying in the boundary of an equilateral, flat $4$-simplex. Such a value of the dihedral angle is also responsible for the mechanism of coherent cancellation of phases, which yields the correct semiclassical behavior of the 2-point function.

In cosmology, de Sitter space-time is usually coordinatized in the form
\begin{align}
ds^2=-dt^2+e^{2Ht}(dx^2+dy^2+dz^2),
\end{align}
so that the constant-$t$ surfaces are flat Euclidean spaces $\mathbb E^3$, and the scale factor grows exponentially in time.  $H$ is the (constant) Hubble rate of expansion. We are not considering here such a kind of canonical surfaces.  We rather consider a spherical slicing of de Sitter space-time attained by the use of the following coordinates:
\begin{align}
ds^2=-dt^2+\frac{1}{H^2}\cosh^2(H t)d\Omega_3
\end{align}
in the Lorentzian case, and
\begin{align}
ds^2=dr^2+\frac{1}{H^2}\cos^2(H r)d\Omega_3
\end{align}
in the Riemannian case. When de Sitter space-time is viewed as the homogeneous, isotropic solution of vacuum Einstein equations with cosmological constant $\Lambda$, we have $H=\sqrt{\Lambda/3}$. This foliation of the de Sitter manifold corresponds, for the two space-time signatures, to
\begin{align}
&K=\dot a(t)=\sinh(Ht)\\
&K=\dot a(r)=-\sin(Hr)
\end{align}
respectively.

As a final remark, notice that if we invert the sign of $K$ in \eqref{asymptotic state}, we obtain the the complex conjugate state, which is a different state, even though classically these two states correspond to the same solution of Einstein equations, but opposite space-time orientations. A different way to think about it is to consider parity transformations, enlarging $SO(3)$ to the full orthogonal group $O(3)$. Under a parity transformation, which is a large gauge transformation, the triad changes sign and the scalar coefficients transform as
\begin{align}
&\Gamma\rightarrow\Gamma,\\
&K\rightarrow -K,
\end{align}
so \eqref{asymptotic state} and its complex conjugate are related by parity. This does not mean that Loop Quantum Gravity violates parity, as parity-related sectors in the kinematical Hilbert space could be super-selected by the dynamics \cite{Bojowald:2007nu,Bojowald:2008nz}. Nevertheless, the issue of the parity behavior of Loop Quantum Gravity is tricky, as one of the fundamental variables, the Ashtekar-Barbero connection, does not transform simply. Moreover, the parity transformation $A_a^i=\Gamma_a^i+\gamma K_a^i\rightarrow \Gamma_a^i-\gamma K_a^i$ requires to disentangle the extrinsic and intrinsic components from the connection, which is possible only by using the equations of motion.
\section*{Conclusions and outlook}
We provided a class of coherent spin-network states for Loop Quantum Gravity which are peaked around $k=1$ FLRW-like geometries. The main result is the derivation of the semiclassical state for flat space-time used in Spin Foams from the canonical theory, and its generalization to curved (homogeneous and isotropic) backgrounds, for both Euclidean and Lorentzian signatures.

Our analysis gives further intuition on which aspects of classical General Relativity are captured by the truncation to a given graph of the phase space of Loop Quantum Gravity. We chose the complete 5-vertex graph (4-simplex graph), symmetrically embedded in the canonical hyper-surface, in order to compare the result with the standard boundary states of Spin Foam vertex amplitudes, but we stress that the construction can be easily generalized to different (e.g. very fine) graphs. 
The applications of such a class of coherent states in a context of cosmological interest could open new perspectives within the semiclassical analysis of the Spin Foam dynamics, as a possible development of a cosmological perturbation theory. 

Taking into account a simple and highly symmetric semiclassical state is the natural way to perform a symmetric reduction within the full quantum theory, which can then be compared with the standard results of Loop Quantum Cosmology. We hope the simple coherent state discussed here (maybe the most simple) could shed light on this relationship. Finally, it would be interesting to investigate the relation (if any) of the de Sitter coherent state we presented in this paper with the Kodama ground state \cite{Kodama:1988,Smolin:2002sz,Freidel:2003pu}.
\section*{Acknowlegements}
We thank Stephon Alexander and Carlo Rovelli for comments on the manuscript. This work was supported in part by the NSF grant PHY0854743, The George A. and Margaret M.~Downsbrough Endowment and the Eberly research funds of Penn State. E.M. gratefully acknowledges support from ``Fondazione Angelo della Riccia''. A.M. acknowledges support from NSF CAREER grant.
\appendix
\section{Useful formulae}
The Maurer-Cartan form $\omega=\omega_a^i \tau^i dx^a$ in hyperspheric coordinates reads
\begin{align}
&\omega^1=\cos\phi \sin\theta d\psi+\sin\psi \big(\cos\psi\cos\phi\cos\theta+\\\nonumber
 &-\sin\psi\sin\phi \big) d\theta
-\sin\psi\sin\theta \big( \sin\psi\cos\phi\cos\theta +\\\nonumber
&+\cos\psi\sin\phi \big) d\phi,\\
&\omega^2=\sin\phi \sin\theta d\psi+\sin\psi \big( \cos\psi\sin\phi\cos\theta+\\\nonumber
 &+\sin\psi\cos\phi \big) d\theta
-\sin\psi\sin\theta \big( \sin\psi\sin\phi\cos\theta +\\ \nonumber
&-\cos\psi\cos\phi \big) d\phi,\\
&\omega^3=\cos\theta d\psi-\sin\psi\cos\psi\sin\theta d\theta
+\sin^2\psi\sin^2\theta d\phi.
\end{align}
with ranges $\theta\in [0,\pi),\phi\in[0,2\pi),\psi\in[0,2\pi)$.
\bibliography{biblioFLRW}

\begin{thebibliography}{60}%
\makeatletter
\providecommand \@ifxundefined [1]{%
 \@ifx{#1\undefined}
}%
\providecommand \@ifnum [1]{%
 \ifnum #1\expandafter \@firstoftwo
 \else \expandafter \@secondoftwo
 \fi
}%
\providecommand \@ifx [1]{%
 \ifx #1\expandafter \@firstoftwo
 \else \expandafter \@secondoftwo
 \fi
}%
\providecommand \natexlab [1]{#1}%
\providecommand \enquote  [1]{``#1''}%
\providecommand \bibnamefont  [1]{#1}%
\providecommand \bibfnamefont [1]{#1}%
\providecommand \citenamefont [1]{#1}%
\providecommand \href@noop [0]{\@secondoftwo}%
\providecommand \href [0]{\begingroup \@sanitize@url \@href}%
\providecommand \@href[1]{\@@startlink{#1}\@@href}%
\providecommand \@@href[1]{\endgroup#1\@@endlink}%
\providecommand \@sanitize@url [0]{\catcode `\\12\catcode `\$12\catcode
  `\&12\catcode `\#12\catcode `\^12\catcode `\_12\catcode `\%12\relax}%
\providecommand \@@startlink[1]{}%
\providecommand \@@endlink[0]{}%
\providecommand \url  [0]{\begingroup\@sanitize@url \@url }%
\providecommand \@url [1]{\endgroup\@href {#1}{\urlprefix }}%
\providecommand \urlprefix  [0]{URL }%
\providecommand \Eprint [0]{\href }%
\@ifxundefined \urlstyle {%
  \providecommand \doi  [0]{\begingroup \@sanitize@url \@doi}%
  \providecommand \@doi [1]{\endgroup \@@startlink {\doibase
  #1}doi:\discretionary {}{}{}#1\@@endlink }%
}{%
  \providecommand \doi  [0]{doi:\discretionary{}{}{}\begingroup
  \urlstyle{rm}\Url }%
}%
\providecommand \doibase [0]{http://dx.doi.org/}%
\providecommand \Doi [0]{\begingroup \@sanitize@url \@Doi }%
\providecommand \@Doi  [1]{\endgroup\@@startlink{\doibase#1}\@@Doi}%
\providecommand \@@Doi [1]{#1\@@endlink}%
\providecommand \selectlanguage [0]{\@gobble}%
\providecommand \bibinfo  [0]{\@secondoftwo}%
\providecommand \bibfield  [0]{\@secondoftwo}%
\providecommand \translation [1]{[#1]}%
\providecommand \BibitemOpen [0]{}%
\providecommand \bibitemStop [0]{}%
\providecommand \bibitemNoStop [0]{.\EOS\space}%
\providecommand \EOS [0]{\spacefactor3000\relax}%
\providecommand \BibitemShut  [1]{\csname bibitem#1\endcsname}%
\bibitem [{\citenamefont {Ashtekar}(1986)}]{Ashtekar:1986yd}%
  \BibitemOpen
  \bibfield  {author} {\bibinfo {author} {\bibfnamefont {Abhay}\ \bibnamefont
  {Ashtekar}},\ }\bibfield  {title} {\enquote {\bibinfo {title} {{New Variables
  for Classical and Quantum Gravity}},}\ }\Doi {10.1103/PhysRevLett.57.2244}
  {\bibfield  {journal} {\bibinfo  {journal} {Phys. Rev. Lett.},\ }\textbf
  {\bibinfo {volume} {57}},\ \bibinfo {pages} {2244--2247} (\bibinfo {year}
  {1986})}\BibitemShut {NoStop}%
\bibitem [{\citenamefont {Rovelli}\ and\ \citenamefont
  {Smolin}(1990)}]{Rovelli:1989za}%
  \BibitemOpen
  \bibfield  {author} {\bibinfo {author} {\bibfnamefont {Carlo}\ \bibnamefont
  {Rovelli}}\ and\ \bibinfo {author} {\bibfnamefont {Lee}\ \bibnamefont
  {Smolin}},\ }\bibfield  {title} {\enquote {\bibinfo {title} {{Loop Space
  Representation of Quantum General Relativity}},}\ }\Doi
  {10.1016/0550-3213(90)90019-A} {\bibfield  {journal} {\bibinfo  {journal}
  {Nucl. Phys.},\ }\textbf {\bibinfo {volume} {B331}},\ \bibinfo {pages} {80}
  (\bibinfo {year} {1990})}\BibitemShut {NoStop}%
\bibitem [{\citenamefont {Ashtekar}\ and\ \citenamefont
  {Isham}(1992)}]{Ashtekar:1991kc}%
  \BibitemOpen
  \bibfield  {author} {\bibinfo {author} {\bibfnamefont {Abhay}\ \bibnamefont
  {Ashtekar}}\ and\ \bibinfo {author} {\bibfnamefont {C.~J.}\ \bibnamefont
  {Isham}},\ }\bibfield  {title} {\enquote {\bibinfo {title} {{Representations
  of the holonomy algebras of gravity and nonAbelian gauge theories}},}\ }\Doi
  {10.1088/0264-9381/9/6/004} {\bibfield  {journal} {\bibinfo  {journal}
  {Class. Quant. Grav.},\ }\textbf {\bibinfo {volume} {9}},\ \bibinfo {pages}
  {1433--1468} (\bibinfo {year} {1992})},\ \Eprint
  {http://arxiv.org/abs/hep-th/9202053} {arXiv:hep-th/9202053} \BibitemShut
  {NoStop}%
\bibitem [{\citenamefont {Rovelli}()}]{Rovelli:2004tv}%
  \BibitemOpen
  \bibfield  {author} {\bibinfo {author} {\bibfnamefont {Carlo}\ \bibnamefont
  {Rovelli}},\ }\bibfield  {title} {\enquote {\bibinfo {title} {{Quantum
  gravity}},}\ }\href@noop {} {}\bibinfo {note} {Cambridge, UK: Univ. Pr.
  (2004) 455 p}\BibitemShut {NoStop}%
\bibitem [{\citenamefont {Thiemann}(2001){\natexlab{a}}}]{Thiemann:2007zz}%
  \BibitemOpen
  \bibfield  {author} {\bibinfo {author} {\bibfnamefont {Thomas}\ \bibnamefont
  {Thiemann}},\ }\bibfield  {title} {\enquote {\bibinfo {title} {{Modern
  canonical quantum general relativity}},}\ }\href@noop {} { (\bibinfo {year}
  {2001}{\natexlab{a}})},\ \Eprint {http://arxiv.org/abs/gr-qc/0110034}
  {arXiv:gr-qc/0110034} \BibitemShut {NoStop}%
\bibitem [{\citenamefont {Bojowald}(2003)}]{Bojowald:2003md}%
  \BibitemOpen
  \bibfield  {author} {\bibinfo {author} {\bibfnamefont {Martin}\ \bibnamefont
  {Bojowald}},\ }\bibfield  {title} {\enquote {\bibinfo {title} {{Homogeneous
  loop quantum cosmology}},}\ }\Doi {10.1088/0264-9381/20/13/310} {\bibfield
  {journal} {\bibinfo  {journal} {Class. Quant. Grav.},\ }\textbf {\bibinfo
  {volume} {20}},\ \bibinfo {pages} {2595--2615} (\bibinfo {year} {2003})},\
  \Eprint {http://arxiv.org/abs/gr-qc/0303073} {arXiv:gr-qc/0303073}
  \BibitemShut {NoStop}%
\bibitem [{\citenamefont {Ashtekar}\ \emph {et~al.}(2003)\citenamefont
  {Ashtekar}, \citenamefont {Bojowald},\ and\ \citenamefont
  {Lewandowski}}]{Ashtekar:2003hd}%
  \BibitemOpen
  \bibfield  {author} {\bibinfo {author} {\bibfnamefont {Abhay}\ \bibnamefont
  {Ashtekar}}, \bibinfo {author} {\bibfnamefont {Martin}\ \bibnamefont
  {Bojowald}}, \ and\ \bibinfo {author} {\bibfnamefont {Jerzy}\ \bibnamefont
  {Lewandowski}},\ }\bibfield  {title} {\enquote {\bibinfo {title}
  {{Mathematical structure of loop quantum cosmology}},}\ }\href@noop {}
  {\bibfield  {journal} {\bibinfo  {journal} {Adv. Theor. Math. Phys.},\
  }\textbf {\bibinfo {volume} {7}},\ \bibinfo {pages} {233--268} (\bibinfo
  {year} {2003})},\ \Eprint {http://arxiv.org/abs/gr-qc/0304074}
  {arXiv:gr-qc/0304074} \BibitemShut {NoStop}%
\bibitem [{\citenamefont {Bojowald}(2001)}]{Bojowald:2001xe}%
  \BibitemOpen
  \bibfield  {author} {\bibinfo {author} {\bibfnamefont {Martin}\ \bibnamefont
  {Bojowald}},\ }\bibfield  {title} {\enquote {\bibinfo {title} {{Absence of
  singularity in loop quantum cosmology}},}\ }\Doi
  {10.1103/PhysRevLett.86.5227} {\bibfield  {journal} {\bibinfo  {journal}
  {Phys. Rev. Lett.},\ }\textbf {\bibinfo {volume} {86}},\ \bibinfo {pages}
  {5227--5230} (\bibinfo {year} {2001})},\ \Eprint
  {http://arxiv.org/abs/gr-qc/0102069} {arXiv:gr-qc/0102069} \BibitemShut
  {NoStop}%
\bibitem [{\citenamefont {Bojowald}(2002)}]{Bojowald:2002gz}%
  \BibitemOpen
  \bibfield  {author} {\bibinfo {author} {\bibfnamefont {Martin}\ \bibnamefont
  {Bojowald}},\ }\bibfield  {title} {\enquote {\bibinfo {title} {{Isotropic
  loop quantum cosmology}},}\ }\Doi {10.1088/0264-9381/19/10/313} {\bibfield
  {journal} {\bibinfo  {journal} {Class. Quant. Grav.},\ }\textbf {\bibinfo
  {volume} {19}},\ \bibinfo {pages} {2717--2742} (\bibinfo {year} {2002})},\
  \Eprint {http://arxiv.org/abs/gr-qc/0202077} {arXiv:gr-qc/0202077}
  \BibitemShut {NoStop}%
\bibitem [{\citenamefont {Ashtekar}(2007)}]{Ashtekar:2007tv}%
  \BibitemOpen
  \bibfield  {author} {\bibinfo {author} {\bibfnamefont {Abhay}\ \bibnamefont
  {Ashtekar}},\ }\bibfield  {title} {\enquote {\bibinfo {title} {{An
  Introduction to Loop Quantum Gravity Through Cosmology}},}\ }\Doi
  {10.1393/ncb/i2007-10351-5} {\bibfield  {journal} {\bibinfo  {journal} {Nuovo
  Cim.},\ }\textbf {\bibinfo {volume} {B122}},\ \bibinfo {pages} {135--155}
  (\bibinfo {year} {2007})},\ \Eprint {http://arxiv.org/abs/gr-qc/0702030}
  {arXiv:gr-qc/0702030} \BibitemShut {NoStop}%
\bibitem [{\citenamefont {Thiemann}\ and\ \citenamefont
  {Winkler}(2001){\natexlab{a}}}]{Thiemann:2000bx}%
  \BibitemOpen
  \bibfield  {author} {\bibinfo {author} {\bibfnamefont {T.}~\bibnamefont
  {Thiemann}}\ and\ \bibinfo {author} {\bibfnamefont {O.}~\bibnamefont
  {Winkler}},\ }\bibfield  {title} {\enquote {\bibinfo {title} {{Gauge field
  theory coherent states (GCS) III: Ehrenfest theorems}},}\ }\Doi
  {10.1088/0264-9381/18/21/315} {\bibfield  {journal} {\bibinfo  {journal}
  {Class. Quant. Grav.},\ }\textbf {\bibinfo {volume} {18}},\ \bibinfo {pages}
  {4629--4682} (\bibinfo {year} {2001}{\natexlab{a}})},\ \Eprint
  {http://arxiv.org/abs/hep-th/0005234} {arXiv:hep-th/0005234} \BibitemShut
  {NoStop}%
\bibitem [{\citenamefont {Bahr}\ and\ \citenamefont
  {Thiemann}(2007)}]{Bahr:2006hm}%
  \BibitemOpen
  \bibfield  {author} {\bibinfo {author} {\bibfnamefont {Benjamin}\
  \bibnamefont {Bahr}}\ and\ \bibinfo {author} {\bibfnamefont {Thomas}\
  \bibnamefont {Thiemann}},\ }\bibfield  {title} {\enquote {\bibinfo {title}
  {{Approximating the physical inner product of Loop Quantum Cosmology}},}\
  }\Doi {10.1088/0264-9381/24/8/011} {\bibfield  {journal} {\bibinfo  {journal}
  {Class. Quant. Grav.},\ }\textbf {\bibinfo {volume} {24}},\ \bibinfo {pages}
  {2109--2138} (\bibinfo {year} {2007})},\ \Eprint
  {http://arxiv.org/abs/gr-qc/0607075} {arXiv:gr-qc/0607075} \BibitemShut
  {NoStop}%
\bibitem [{\citenamefont {Sahlmann}\ and\ \citenamefont
  {Thiemann}(2006){\natexlab{a}}}]{Sahlmann:2002qj}%
  \BibitemOpen
  \bibfield  {author} {\bibinfo {author} {\bibfnamefont {Hanno}\ \bibnamefont
  {Sahlmann}}\ and\ \bibinfo {author} {\bibfnamefont {Thomas}\ \bibnamefont
  {Thiemann}},\ }\bibfield  {title} {\enquote {\bibinfo {title} {{Towards the
  QFT on curved spacetime limit of QGR. I: A general scheme}},}\ }\Doi
  {10.1088/0264-9381/23/3/019} {\bibfield  {journal} {\bibinfo  {journal}
  {Class. Quant. Grav.},\ }\textbf {\bibinfo {volume} {23}},\ \bibinfo {pages}
  {867--908} (\bibinfo {year} {2006}{\natexlab{a}})},\ \Eprint
  {http://arxiv.org/abs/gr-qc/0207030} {arXiv:gr-qc/0207030} \BibitemShut
  {NoStop}%
\bibitem [{\citenamefont {Sahlmann}\ and\ \citenamefont
  {Thiemann}(2006){\natexlab{b}}}]{Sahlmann:2002qk}%
  \BibitemOpen
  \bibfield  {author} {\bibinfo {author} {\bibfnamefont {Hanno}\ \bibnamefont
  {Sahlmann}}\ and\ \bibinfo {author} {\bibfnamefont {Thomas}\ \bibnamefont
  {Thiemann}},\ }\bibfield  {title} {\enquote {\bibinfo {title} {{Towards the
  QFT on curved spacetime limit of QGR. II: A concrete implementation}},}\
  }\Doi {10.1088/0264-9381/23/3/020} {\bibfield  {journal} {\bibinfo  {journal}
  {Class. Quant. Grav.},\ }\textbf {\bibinfo {volume} {23}},\ \bibinfo {pages}
  {909--954} (\bibinfo {year} {2006}{\natexlab{b}})},\ \Eprint
  {http://arxiv.org/abs/gr-qc/0207031} {arXiv:gr-qc/0207031} \BibitemShut
  {NoStop}%
\bibitem [{\citenamefont {Reisenberger}\ and\ \citenamefont
  {Rovelli}(2000)}]{Reisenberger:2000fy}%
  \BibitemOpen
  \bibfield  {author} {\bibinfo {author} {\bibfnamefont {Michael}\ \bibnamefont
  {Reisenberger}}\ and\ \bibinfo {author} {\bibfnamefont {Carlo}\ \bibnamefont
  {Rovelli}},\ }\bibfield  {title} {\enquote {\bibinfo {title} {{Spin foams as
  Feynman diagrams}},}\ }\href@noop {} { (\bibinfo {year} {2000})},\ \Eprint
  {http://arxiv.org/abs/gr-qc/0002083} {arXiv:gr-qc/0002083} \BibitemShut
  {NoStop}%
\bibitem [{\citenamefont {Perez}(2003)}]{Perez:2003vx}%
  \BibitemOpen
  \bibfield  {author} {\bibinfo {author} {\bibfnamefont {Alejandro}\
  \bibnamefont {Perez}},\ }\bibfield  {title} {\enquote {\bibinfo {title}
  {{Spin foam models for quantum gravity}},}\ }\href@noop {} {\bibfield
  {journal} {\bibinfo  {journal} {Class. Quant. Grav.},\ }\textbf {\bibinfo
  {volume} {20}},\ \bibinfo {pages} {R43} (\bibinfo {year} {2003})},\ \Eprint
  {http://arxiv.org/abs/gr-qc/0301113} {arXiv:gr-qc/0301113} \BibitemShut
  {NoStop}%
\bibitem [{\citenamefont {Ashtekar}\ \emph {et~al.}(2010)\citenamefont
  {Ashtekar}, \citenamefont {Campiglia},\ and\ \citenamefont
  {Henderson}}]{Ashtekar:2010ve}%
  \BibitemOpen
  \bibfield  {author} {\bibinfo {author} {\bibfnamefont {Abhay}\ \bibnamefont
  {Ashtekar}}, \bibinfo {author} {\bibfnamefont {Miguel}\ \bibnamefont
  {Campiglia}}, \ and\ \bibinfo {author} {\bibfnamefont {Adam}\ \bibnamefont
  {Henderson}},\ }\bibfield  {title} {\enquote {\bibinfo {title} {{Casting Loop
  Quantum Cosmology in the Spin Foam Paradigm}},}\ }\Doi
  {10.1088/0264-9381/27/13/135020} {\bibfield  {journal} {\bibinfo  {journal}
  {Class. Quant. Grav.},\ }\textbf {\bibinfo {volume} {27}},\ \bibinfo {pages}
  {135020} (\bibinfo {year} {2010})},\ \Eprint {http://arxiv.org/abs/1001.5147}
  {arXiv:1001.5147 [gr-qc]} \BibitemShut {NoStop}%
\bibitem [{\citenamefont {Livine}\ and\ \citenamefont
  {Speziale}(2007)}]{Livine:2007vk}%
  \BibitemOpen
  \bibfield  {author} {\bibinfo {author} {\bibfnamefont {Etera~R.}\
  \bibnamefont {Livine}}\ and\ \bibinfo {author} {\bibfnamefont {Simone}\
  \bibnamefont {Speziale}},\ }\bibfield  {title} {\enquote {\bibinfo {title}
  {{A new spinfoam vertex for quantum gravity}},}\ }\Doi
  {10.1103/PhysRevD.76.084028} {\bibfield  {journal} {\bibinfo  {journal}
  {Phys. Rev.},\ }\textbf {\bibinfo {volume} {D76}},\ \bibinfo {pages} {084028}
  (\bibinfo {year} {2007})},\ \Eprint {http://arxiv.org/abs/0705.0674}
  {arXiv:0705.0674 [gr-qc]} \BibitemShut {NoStop}%
\bibitem [{\citenamefont {Engle}\ \emph {et~al.}(2008)\citenamefont {Engle},
  \citenamefont {Livine}, \citenamefont {Pereira},\ and\ \citenamefont
  {Rovelli}}]{Engle:2007wy}%
  \BibitemOpen
  \bibfield  {author} {\bibinfo {author} {\bibfnamefont {Jonathan}\
  \bibnamefont {Engle}}, \bibinfo {author} {\bibfnamefont {Etera}\ \bibnamefont
  {Livine}}, \bibinfo {author} {\bibfnamefont {Roberto}\ \bibnamefont
  {Pereira}}, \ and\ \bibinfo {author} {\bibfnamefont {Carlo}\ \bibnamefont
  {Rovelli}},\ }\bibfield  {title} {\enquote {\bibinfo {title} {{LQG vertex
  with finite Immirzi parameter}},}\ }\Doi {10.1016/j.nuclphysb.2008.02.018}
  {\bibfield  {journal} {\bibinfo  {journal} {Nucl. Phys.},\ }\textbf {\bibinfo
  {volume} {B799}},\ \bibinfo {pages} {136--149} (\bibinfo {year} {2008})},\
  \Eprint {http://arxiv.org/abs/0711.0146} {arXiv:0711.0146 [gr-qc]}
  \BibitemShut {NoStop}%
\bibitem [{\citenamefont {Freidel}\ and\ \citenamefont
  {Krasnov}(2008)}]{Freidel:2007py}%
  \BibitemOpen
  \bibfield  {author} {\bibinfo {author} {\bibfnamefont {Laurent}\ \bibnamefont
  {Freidel}}\ and\ \bibinfo {author} {\bibfnamefont {Kirill}\ \bibnamefont
  {Krasnov}},\ }\bibfield  {title} {\enquote {\bibinfo {title} {{A New Spin
  Foam Model for 4d Gravity}},}\ }\Doi {10.1088/0264-9381/25/12/125018}
  {\bibfield  {journal} {\bibinfo  {journal} {Class. Quant. Grav.},\ }\textbf
  {\bibinfo {volume} {25}},\ \bibinfo {pages} {125018} (\bibinfo {year}
  {2008})},\ \Eprint {http://arxiv.org/abs/0708.1595} {arXiv:0708.1595 [gr-qc]}
  \BibitemShut {NoStop}%
\bibitem [{\citenamefont {Magliaro}\ \emph {et~al.}(2008)\citenamefont
  {Magliaro}, \citenamefont {Perini},\ and\ \citenamefont
  {Rovelli}}]{Magliaro:2007nc}%
  \BibitemOpen
  \bibfield  {author} {\bibinfo {author} {\bibfnamefont {Elena}\ \bibnamefont
  {Magliaro}}, \bibinfo {author} {\bibfnamefont {Claudio}\ \bibnamefont
  {Perini}}, \ and\ \bibinfo {author} {\bibfnamefont {Carlo}\ \bibnamefont
  {Rovelli}},\ }\bibfield  {title} {\enquote {\bibinfo {title} {{Numerical
  indications on the semiclassical limit of the flipped vertex}},}\ }\Doi
  {10.1088/0264-9381/25/9/095009} {\bibfield  {journal} {\bibinfo  {journal}
  {Class. Quant. Grav.},\ }\textbf {\bibinfo {volume} {25}},\ \bibinfo {pages}
  {095009} (\bibinfo {year} {2008})},\ \Eprint {http://arxiv.org/abs/0710.5034}
  {arXiv:0710.5034 [gr-qc]} \BibitemShut {NoStop}%
\bibitem [{\citenamefont {Alesci}\ \emph {et~al.}(2009)\citenamefont {Alesci},
  \citenamefont {Bianchi}, \citenamefont {Magliaro},\ and\ \citenamefont
  {Perini}}]{Alesci:2008ec}%
  \BibitemOpen
  \bibfield  {author} {\bibinfo {author} {\bibfnamefont {Emanuele}\
  \bibnamefont {Alesci}}, \bibinfo {author} {\bibfnamefont {Eugenio}\
  \bibnamefont {Bianchi}}, \bibinfo {author} {\bibfnamefont {Elena}\
  \bibnamefont {Magliaro}}, \ and\ \bibinfo {author} {\bibfnamefont {Claudio}\
  \bibnamefont {Perini}},\ }\bibfield  {title} {\enquote {\bibinfo {title}
  {{Intertwiner dynamics in the flipped vertex}},}\ }\Doi
  {10.1088/0264-9381/26/18/185003} {\bibfield  {journal} {\bibinfo  {journal}
  {Class. Quant. Grav.},\ }\textbf {\bibinfo {volume} {26}},\ \bibinfo {pages}
  {185003} (\bibinfo {year} {2009})},\ \Eprint {http://arxiv.org/abs/0808.1971}
  {arXiv:0808.1971 [gr-qc]} \BibitemShut {NoStop}%
\bibitem [{\citenamefont {Alesci}\ \emph {et~al.}(2010)\citenamefont {Alesci},
  \citenamefont {Bianchi}, \citenamefont {Magliaro},\ and\ \citenamefont
  {Perini}}]{Alesci:2008un}%
  \BibitemOpen
  \bibfield  {author} {\bibinfo {author} {\bibfnamefont {Emanuele}\
  \bibnamefont {Alesci}}, \bibinfo {author} {\bibfnamefont {Eugenio}\
  \bibnamefont {Bianchi}}, \bibinfo {author} {\bibfnamefont {Elena}\
  \bibnamefont {Magliaro}}, \ and\ \bibinfo {author} {\bibfnamefont {Claudio}\
  \bibnamefont {Perini}},\ }\bibfield  {title} {\enquote {\bibinfo {title}
  {{Asymptotics of LQG fusion coefficients}},}\ }\Doi
  {10.1088/0264-9381/27/9/095016} {\bibfield  {journal} {\bibinfo  {journal}
  {Class. Quant. Grav.},\ }\textbf {\bibinfo {volume} {27}},\ \bibinfo {pages}
  {095016} (\bibinfo {year} {2010})},\ \Eprint {http://arxiv.org/abs/0809.3718}
  {arXiv:0809.3718 [gr-qc]} \BibitemShut {NoStop}%
\bibitem [{\citenamefont {Conrady}\ and\ \citenamefont
  {Freidel}(2008)}]{Conrady:2008mk}%
  \BibitemOpen
  \bibfield  {author} {\bibinfo {author} {\bibfnamefont {Florian}\ \bibnamefont
  {Conrady}}\ and\ \bibinfo {author} {\bibfnamefont {Laurent}\ \bibnamefont
  {Freidel}},\ }\bibfield  {title} {\enquote {\bibinfo {title} {{On the
  semiclassical limit of 4d spin foam models}},}\ }\Doi
  {10.1103/PhysRevD.78.104023} {\bibfield  {journal} {\bibinfo  {journal}
  {Phys. Rev.},\ }\textbf {\bibinfo {volume} {D78}},\ \bibinfo {pages} {104023}
  (\bibinfo {year} {2008})},\ \Eprint {http://arxiv.org/abs/0809.2280}
  {arXiv:0809.2280 [gr-qc]} \BibitemShut {NoStop}%
\bibitem [{\citenamefont {Barrett}\ \emph {et~al.}(2009)\citenamefont
  {Barrett}, \citenamefont {Dowdall}, \citenamefont {Fairbairn}, \citenamefont
  {Gomes},\ and\ \citenamefont {Hellmann}}]{Barrett:2009gg}%
  \BibitemOpen
  \bibfield  {author} {\bibinfo {author} {\bibfnamefont {John~W.}\ \bibnamefont
  {Barrett}}, \bibinfo {author} {\bibfnamefont {Richard~J.}\ \bibnamefont
  {Dowdall}}, \bibinfo {author} {\bibfnamefont {Winston~J.}\ \bibnamefont
  {Fairbairn}}, \bibinfo {author} {\bibfnamefont {Henrique}\ \bibnamefont
  {Gomes}}, \ and\ \bibinfo {author} {\bibfnamefont {Frank}\ \bibnamefont
  {Hellmann}},\ }\bibfield  {title} {\enquote {\bibinfo {title} {{Asymptotic
  analysis of the EPRL four-simplex amplitude}},}\ }\Doi {10.1063/1.3244218}
  {\bibfield  {journal} {\bibinfo  {journal} {J. Math. Phys.},\ }\textbf
  {\bibinfo {volume} {50}},\ \bibinfo {pages} {112504} (\bibinfo {year}
  {2009})},\ \Eprint {http://arxiv.org/abs/0902.1170} {arXiv:0902.1170 [gr-qc]}
  \BibitemShut {NoStop}%
\bibitem [{\citenamefont {Barrett}\ \emph {et~al.}(2010)\citenamefont
  {Barrett}, \citenamefont {Dowdall}, \citenamefont {Fairbairn}, \citenamefont
  {Hellmann},\ and\ \citenamefont {Pereira}}]{Barrett:2009mw}%
  \BibitemOpen
  \bibfield  {author} {\bibinfo {author} {\bibfnamefont {John~W.}\ \bibnamefont
  {Barrett}}, \bibinfo {author} {\bibfnamefont {Richard~J.}\ \bibnamefont
  {Dowdall}}, \bibinfo {author} {\bibfnamefont {Winston~J.}\ \bibnamefont
  {Fairbairn}}, \bibinfo {author} {\bibfnamefont {Frank}\ \bibnamefont
  {Hellmann}}, \ and\ \bibinfo {author} {\bibfnamefont {Roberto}\ \bibnamefont
  {Pereira}},\ }\bibfield  {title} {\enquote {\bibinfo {title} {{Lorentzian
  spin foam amplitudes: graphical calculus and asymptotics}},}\ }\Doi
  {10.1088/0264-9381/27/16/165009} {\bibfield  {journal} {\bibinfo  {journal}
  {Class. Quant. Grav.},\ }\textbf {\bibinfo {volume} {27}},\ \bibinfo {pages}
  {165009} (\bibinfo {year} {2010})},\ \Eprint {http://arxiv.org/abs/0907.2440}
  {arXiv:0907.2440 [gr-qc]} \BibitemShut {NoStop}%
\bibitem [{\citenamefont {Perini}\ \emph {et~al.}(2009)\citenamefont {Perini},
  \citenamefont {Rovelli},\ and\ \citenamefont {Speziale}}]{Perini:2008pd}%
  \BibitemOpen
  \bibfield  {author} {\bibinfo {author} {\bibfnamefont {Claudio}\ \bibnamefont
  {Perini}}, \bibinfo {author} {\bibfnamefont {Carlo}\ \bibnamefont {Rovelli}},
  \ and\ \bibinfo {author} {\bibfnamefont {Simone}\ \bibnamefont {Speziale}},\
  }\bibfield  {title} {\enquote {\bibinfo {title} {{Self-energy and vertex
  radiative corrections in LQG}},}\ }\Doi {10.1016/j.physletb.2009.10.076}
  {\bibfield  {journal} {\bibinfo  {journal} {Phys. Lett.},\ }\textbf {\bibinfo
  {volume} {B682}},\ \bibinfo {pages} {78--84} (\bibinfo {year} {2009})},\
  \Eprint {http://arxiv.org/abs/0810.1714} {arXiv:0810.1714 [gr-qc]}
  \BibitemShut {NoStop}%
\bibitem [{\citenamefont {Krajewski}\ \emph {et~al.}(2010)\citenamefont
  {Krajewski}, \citenamefont {Magnen}, \citenamefont {Rivasseau}, \citenamefont
  {Tanasa},\ and\ \citenamefont {Vitale}}]{Krajewski:2010yq}%
  \BibitemOpen
  \bibfield  {author} {\bibinfo {author} {\bibfnamefont {Thomas}\ \bibnamefont
  {Krajewski}}, \bibinfo {author} {\bibfnamefont {Jacques}\ \bibnamefont
  {Magnen}}, \bibinfo {author} {\bibfnamefont {Vincent}\ \bibnamefont
  {Rivasseau}}, \bibinfo {author} {\bibfnamefont {Adrian}\ \bibnamefont
  {Tanasa}}, \ and\ \bibinfo {author} {\bibfnamefont {Patrizia}\ \bibnamefont
  {Vitale}},\ }\bibfield  {title} {\enquote {\bibinfo {title} {{Quantum
  Corrections in the Group Field Theory Formulation of the EPRL/FK Models}},}\
  }\href@noop {} { (\bibinfo {year} {2010})},\ \Eprint
  {http://arxiv.org/abs/1007.3150} {arXiv:1007.3150 [gr-qc]} \BibitemShut
  {NoStop}%
\bibitem [{\citenamefont {Geloun}\ \emph {et~al.}(2010)\citenamefont {Geloun},
  \citenamefont {Gurau},\ and\ \citenamefont {Rivasseau}}]{Geloun:2010vj}%
  \BibitemOpen
  \bibfield  {author} {\bibinfo {author} {\bibfnamefont {Joseph~Ben}\
  \bibnamefont {Geloun}}, \bibinfo {author} {\bibfnamefont {Razvan}\
  \bibnamefont {Gurau}}, \ and\ \bibinfo {author} {\bibfnamefont {Vincent}\
  \bibnamefont {Rivasseau}},\ }\bibfield  {title} {\enquote {\bibinfo {title}
  {{EPRL/FK Group Field Theory}},}\ }\href@noop {} { (\bibinfo {year}
  {2010})},\ \Eprint {http://arxiv.org/abs/1008.0354} {arXiv:1008.0354
  [hep-th]} \BibitemShut {NoStop}%
\bibitem [{\citenamefont {Oeckl}(2003)}]{Oeckl:2003vu}%
  \BibitemOpen
  \bibfield  {author} {\bibinfo {author} {\bibfnamefont {Robert}\ \bibnamefont
  {Oeckl}},\ }\bibfield  {title} {\enquote {\bibinfo {title} {{A 'general
  boundary' formulation for quantum mechanics and quantum gravity}},}\ }\Doi
  {10.1016/j.physletb.2003.08.043} {\bibfield  {journal} {\bibinfo  {journal}
  {Phys. Lett.},\ }\textbf {\bibinfo {volume} {B575}},\ \bibinfo {pages}
  {318--324} (\bibinfo {year} {2003})},\ \Eprint
  {http://arxiv.org/abs/hep-th/0306025} {arXiv:hep-th/0306025} \BibitemShut
  {NoStop}%
\bibitem [{\citenamefont {Rovelli}(2006)}]{Rovelli:2005yj}%
  \BibitemOpen
  \bibfield  {author} {\bibinfo {author} {\bibfnamefont {Carlo}\ \bibnamefont
  {Rovelli}},\ }\bibfield  {title} {\enquote {\bibinfo {title} {{Graviton
  propagator from background-independent quantum gravity}},}\ }\Doi
  {10.1103/PhysRevLett.97.151301} {\bibfield  {journal} {\bibinfo  {journal}
  {Phys. Rev. Lett.},\ }\textbf {\bibinfo {volume} {97}},\ \bibinfo {pages}
  {151301} (\bibinfo {year} {2006})},\ \Eprint
  {http://arxiv.org/abs/gr-qc/0508124} {arXiv:gr-qc/0508124} \BibitemShut
  {NoStop}%
\bibitem [{\citenamefont {Modesto}\ and\ \citenamefont
  {Rovelli}(2005)}]{Modesto:2005sj}%
  \BibitemOpen
  \bibfield  {author} {\bibinfo {author} {\bibfnamefont {Leonardo}\
  \bibnamefont {Modesto}}\ and\ \bibinfo {author} {\bibfnamefont {Carlo}\
  \bibnamefont {Rovelli}},\ }\bibfield  {title} {\enquote {\bibinfo {title}
  {{Particle scattering in loop quantum gravity}},}\ }\Doi
  {10.1103/PhysRevLett.95.191301} {\bibfield  {journal} {\bibinfo  {journal}
  {Phys. Rev. Lett.},\ }\textbf {\bibinfo {volume} {95}},\ \bibinfo {pages}
  {191301} (\bibinfo {year} {2005})},\ \Eprint
  {http://arxiv.org/abs/gr-qc/0502036} {arXiv:gr-qc/0502036} \BibitemShut
  {NoStop}%
\bibitem [{\citenamefont {Bianchi}\ \emph {et~al.}(2006)\citenamefont
  {Bianchi}, \citenamefont {Modesto}, \citenamefont {Rovelli},\ and\
  \citenamefont {Speziale}}]{Bianchi:2006uf}%
  \BibitemOpen
  \bibfield  {author} {\bibinfo {author} {\bibfnamefont {Eugenio}\ \bibnamefont
  {Bianchi}}, \bibinfo {author} {\bibfnamefont {Leonardo}\ \bibnamefont
  {Modesto}}, \bibinfo {author} {\bibfnamefont {Carlo}\ \bibnamefont
  {Rovelli}}, \ and\ \bibinfo {author} {\bibfnamefont {Simone}\ \bibnamefont
  {Speziale}},\ }\bibfield  {title} {\enquote {\bibinfo {title} {{Graviton
  propagator in loop quantum gravity}},}\ }\Doi {10.1088/0264-9381/23/23/024}
  {\bibfield  {journal} {\bibinfo  {journal} {Class. Quant. Grav.},\ }\textbf
  {\bibinfo {volume} {23}},\ \bibinfo {pages} {6989--7028} (\bibinfo {year}
  {2006})},\ \Eprint {http://arxiv.org/abs/gr-qc/0604044} {arXiv:gr-qc/0604044}
  \BibitemShut {NoStop}%
\bibitem [{\citenamefont {Alesci}\ and\ \citenamefont
  {Rovelli}(2007)}]{Alesci:2007tx}%
  \BibitemOpen
  \bibfield  {author} {\bibinfo {author} {\bibfnamefont {Emanuele}\
  \bibnamefont {Alesci}}\ and\ \bibinfo {author} {\bibfnamefont {Carlo}\
  \bibnamefont {Rovelli}},\ }\bibfield  {title} {\enquote {\bibinfo {title}
  {{The complete LQG propagator: I. Difficulties with the Barrett-Crane
  vertex}},}\ }\Doi {10.1103/PhysRevD.76.104012} {\bibfield  {journal}
  {\bibinfo  {journal} {Phys. Rev.},\ }\textbf {\bibinfo {volume} {D76}},\
  \bibinfo {pages} {104012} (\bibinfo {year} {2007})},\ \Eprint
  {http://arxiv.org/abs/0708.0883} {arXiv:0708.0883 [gr-qc]} \BibitemShut
  {NoStop}%
\bibitem [{\citenamefont {Bianchi}\ \emph {et~al.}(2009)\citenamefont
  {Bianchi}, \citenamefont {Magliaro},\ and\ \citenamefont
  {Perini}}]{Bianchi:2009ri}%
  \BibitemOpen
  \bibfield  {author} {\bibinfo {author} {\bibfnamefont {Eugenio}\ \bibnamefont
  {Bianchi}}, \bibinfo {author} {\bibfnamefont {Elena}\ \bibnamefont
  {Magliaro}}, \ and\ \bibinfo {author} {\bibfnamefont {Claudio}\ \bibnamefont
  {Perini}},\ }\bibfield  {title} {\enquote {\bibinfo {title} {{LQG propagator
  from the new spin foams}},}\ }\Doi {10.1016/j.nuclphysb.2009.07.016}
  {\bibfield  {journal} {\bibinfo  {journal} {Nucl. Phys.},\ }\textbf {\bibinfo
  {volume} {B822}},\ \bibinfo {pages} {245--269} (\bibinfo {year} {2009})},\
  \Eprint {http://arxiv.org/abs/0905.4082} {arXiv:0905.4082 [gr-qc]}
  \BibitemShut {NoStop}%
\bibitem [{\citenamefont {Rovelli}\ and\ \citenamefont
  {Vidotto}(2008)}]{Rovelli:2008dx}%
  \BibitemOpen
  \bibfield  {author} {\bibinfo {author} {\bibfnamefont {Carlo}\ \bibnamefont
  {Rovelli}}\ and\ \bibinfo {author} {\bibfnamefont {Francesca}\ \bibnamefont
  {Vidotto}},\ }\bibfield  {title} {\enquote {\bibinfo {title} {{Stepping out
  of Homogeneity in Loop Quantum Cosmology}},}\ }\Doi
  {10.1088/0264-9381/25/22/225024} {\bibfield  {journal} {\bibinfo  {journal}
  {Class. Quant. Grav.},\ }\textbf {\bibinfo {volume} {25}},\ \bibinfo {pages}
  {225024} (\bibinfo {year} {2008})},\ \Eprint {http://arxiv.org/abs/0805.4585}
  {arXiv:0805.4585 [gr-qc]} \BibitemShut {NoStop}%
\bibitem [{\citenamefont {Battisti}\ \emph {et~al.}(2010)\citenamefont
  {Battisti}, \citenamefont {Marcian\`o},\ and\ \citenamefont
  {Rovelli}}]{Battisti:2009kp}%
  \BibitemOpen
  \bibfield  {author} {\bibinfo {author} {\bibfnamefont {Marco~Valerio}\
  \bibnamefont {Battisti}}, \bibinfo {author} {\bibfnamefont {Antonino}\
  \bibnamefont {Marcian\`o}}, \ and\ \bibinfo {author} {\bibfnamefont {Carlo}\
  \bibnamefont {Rovelli}},\ }\bibfield  {title} {\enquote {\bibinfo {title}
  {{Triangulated Loop Quantum Cosmology: Bianchi IX and inhomogenous
  perturbations}},}\ }\Doi {10.1103/PhysRevD.81.064019} {\bibfield  {journal}
  {\bibinfo  {journal} {Phys. Rev.},\ }\textbf {\bibinfo {volume} {D81}},\
  \bibinfo {pages} {064019} (\bibinfo {year} {2010})},\ \Eprint
  {http://arxiv.org/abs/0911.2653} {arXiv:0911.2653 [gr-qc]} \BibitemShut
  {NoStop}%
\bibitem [{\citenamefont {Marcian\`o}(2010)}]{Marciano:2010jc}%
  \BibitemOpen
  \bibfield  {author} {\bibinfo {author} {\bibfnamefont {Antonino}\
  \bibnamefont {Marcian\`o}},\ }\bibfield  {title} {\enquote {\bibinfo {title}
  {{Towards inhomogeneous loop quantum cosmology: triangulating Bianchi IX with
  perturbations}},}\ }\href@noop {} { (\bibinfo {year} {2010})},\ \Eprint
  {http://arxiv.org/abs/1003.0352} {arXiv:1003.0352 [gr-qc]} \BibitemShut
  {NoStop}%
\bibitem [{\citenamefont {Battisti}\ and\ \citenamefont
  {Marcian\`o}(2010)}]{Battisti:2010he}%
  \BibitemOpen
  \bibfield  {author} {\bibinfo {author} {\bibfnamefont {Marco~Valerio}\
  \bibnamefont {Battisti}}\ and\ \bibinfo {author} {\bibfnamefont {Antonino}\
  \bibnamefont {Marcian\`o}},\ }\bibfield  {title} {\enquote {\bibinfo {title}
  {{Big Bounce in Dipole Cosmology}},}\ }\href@noop {} { (\bibinfo {year}
  {2010})},\ \Eprint {http://arxiv.org/abs/1010.1258} {arXiv:1010.1258 [gr-qc]}
  \BibitemShut {NoStop}%
\bibitem [{\citenamefont {Bianchi}\ \emph
  {et~al.}(2010){\natexlab{a}}\citenamefont {Bianchi}, \citenamefont
  {Rovelli},\ and\ \citenamefont {Vidotto}}]{Bianchi:2010zs}%
  \BibitemOpen
  \bibfield  {author} {\bibinfo {author} {\bibfnamefont {Eugenio}\ \bibnamefont
  {Bianchi}}, \bibinfo {author} {\bibfnamefont {Carlo}\ \bibnamefont
  {Rovelli}}, \ and\ \bibinfo {author} {\bibfnamefont {Francesca}\ \bibnamefont
  {Vidotto}},\ }\bibfield  {title} {\enquote {\bibinfo {title} {{Towards
  Spinfoam Cosmology}},}\ }\Doi {10.1103/PhysRevD.82.084035} {\bibfield
  {journal} {\bibinfo  {journal} {Phys. Rev.},\ }\textbf {\bibinfo {volume}
  {D82}},\ \bibinfo {pages} {084035} (\bibinfo {year} {2010}{\natexlab{a}})},\
  \Eprint {http://arxiv.org/abs/1003.3483} {arXiv:1003.3483 [gr-qc]}
  \BibitemShut {NoStop}%
\bibitem [{\citenamefont {Rovelli}(2010)}]{Rovelli:2010wq}%
  \BibitemOpen
  \bibfield  {author} {\bibinfo {author} {\bibfnamefont {Carlo}\ \bibnamefont
  {Rovelli}},\ }\bibfield  {title} {\enquote {\bibinfo {title} {{A new look at
  loop quantum gravity}},}\ }\href@noop {} { (\bibinfo {year} {2010})},\
  \Eprint {http://arxiv.org/abs/1004.1780} {arXiv:1004.1780 [gr-qc]}
  \BibitemShut {NoStop}%
\bibitem [{\citenamefont {Bianchi}\ \emph
  {et~al.}(2010){\natexlab{b}}\citenamefont {Bianchi}, \citenamefont
  {Magliaro},\ and\ \citenamefont {Perini}}]{Bianchi:2009ky}%
  \BibitemOpen
  \bibfield  {author} {\bibinfo {author} {\bibfnamefont {Eugenio}\ \bibnamefont
  {Bianchi}}, \bibinfo {author} {\bibfnamefont {Elena}\ \bibnamefont
  {Magliaro}}, \ and\ \bibinfo {author} {\bibfnamefont {Claudio}\ \bibnamefont
  {Perini}},\ }\bibfield  {title} {\enquote {\bibinfo {title} {{Coherent
  spin-networks}},}\ }\Doi {10.1103/PhysRevD.82.024012} {\bibfield  {journal}
  {\bibinfo  {journal} {Phys. Rev.},\ }\textbf {\bibinfo {volume} {D82}},\
  \bibinfo {pages} {024012} (\bibinfo {year} {2010}{\natexlab{b}})},\ \Eprint
  {http://arxiv.org/abs/0912.4054} {arXiv:0912.4054 [gr-qc]} \BibitemShut
  {NoStop}%
\bibitem [{\citenamefont {Dasgupta}(2003)}]{Dasgupta:2003da}%
  \BibitemOpen
  \bibfield  {author} {\bibinfo {author} {\bibfnamefont {Arundhati}\
  \bibnamefont {Dasgupta}},\ }\bibfield  {title} {\enquote {\bibinfo {title}
  {{Coherent states for black holes}},}\ }\Doi {10.1088/1475-7516/2003/08/004}
  {\bibfield  {journal} {\bibinfo  {journal} {JCAP},\ }\textbf {\bibinfo
  {volume} {0308}},\ \bibinfo {pages} {004} (\bibinfo {year} {2003})},\ \Eprint
  {http://arxiv.org/abs/hep-th/0305131} {arXiv:hep-th/0305131} \BibitemShut
  {NoStop}%
\bibitem [{\citenamefont {Freidel}\ and\ \citenamefont
  {Speziale}(2010){\natexlab{a}}}]{Freidel:2010aq}%
  \BibitemOpen
  \bibfield  {author} {\bibinfo {author} {\bibfnamefont {Laurent}\ \bibnamefont
  {Freidel}}\ and\ \bibinfo {author} {\bibfnamefont {Simone}\ \bibnamefont
  {Speziale}},\ }\bibfield  {title} {\enquote {\bibinfo {title} {{Twisted
  geometries: A geometric parametrisation of SU(2) phase space}},}\ }\Doi
  {10.1103/PhysRevD.82.084040} {\bibfield  {journal} {\bibinfo  {journal}
  {Phys. Rev.},\ }\textbf {\bibinfo {volume} {D82}},\ \bibinfo {pages} {084040}
  (\bibinfo {year} {2010}{\natexlab{a}})},\ \Eprint
  {http://arxiv.org/abs/1001.2748} {arXiv:1001.2748 [gr-qc]} \BibitemShut
  {NoStop}%
\bibitem [{\citenamefont {Rovelli}\ and\ \citenamefont
  {Speziale}(2010)}]{Rovelli:2010km}%
  \BibitemOpen
  \bibfield  {author} {\bibinfo {author} {\bibfnamefont {Carlo}\ \bibnamefont
  {Rovelli}}\ and\ \bibinfo {author} {\bibfnamefont {Simone}\ \bibnamefont
  {Speziale}},\ }\bibfield  {title} {\enquote {\bibinfo {title} {{On the
  geometry of loop quantum gravity on a graph}},}\ }\Doi
  {10.1103/PhysRevD.82.044018} {\bibfield  {journal} {\bibinfo  {journal}
  {Phys. Rev.},\ }\textbf {\bibinfo {volume} {D82}},\ \bibinfo {pages} {044018}
  (\bibinfo {year} {2010})},\ \Eprint {http://arxiv.org/abs/1005.2927}
  {arXiv:1005.2927 [gr-qc]} \BibitemShut {NoStop}%
\bibitem [{\citenamefont {Freidel}\ and\ \citenamefont
  {Speziale}(2010){\natexlab{b}}}]{Freidel:2010bw}%
  \BibitemOpen
  \bibfield  {author} {\bibinfo {author} {\bibfnamefont {Laurent}\ \bibnamefont
  {Freidel}}\ and\ \bibinfo {author} {\bibfnamefont {Simone}\ \bibnamefont
  {Speziale}},\ }\bibfield  {title} {\enquote {\bibinfo {title} {{From twistors
  to twisted geometries}},}\ }\Doi {10.1103/PhysRevD.82.084041} {\bibfield
  {journal} {\bibinfo  {journal} {Phys. Rev.},\ }\textbf {\bibinfo {volume}
  {D82}},\ \bibinfo {pages} {084041} (\bibinfo {year} {2010}{\natexlab{b}})},\
  \Eprint {http://arxiv.org/abs/1006.0199} {arXiv:1006.0199 [gr-qc]}
  \BibitemShut {NoStop}%
\bibitem [{\citenamefont {Thiemann}(2001){\natexlab{b}}}]{Thiemann:2000bv}%
  \BibitemOpen
  \bibfield  {author} {\bibinfo {author} {\bibfnamefont {T.}~\bibnamefont
  {Thiemann}},\ }\bibfield  {title} {\enquote {\bibinfo {title} {{Quantum spin
  dynamics (QSD). VII: Symplectic structures and continuum lattice formulations
  of gauge field theories}},}\ }\Doi {10.1088/0264-9381/18/17/301} {\bibfield
  {journal} {\bibinfo  {journal} {Class. Quant. Grav.},\ }\textbf {\bibinfo
  {volume} {18}},\ \bibinfo {pages} {3293--3338} (\bibinfo {year}
  {2001}{\natexlab{b}})},\ \Eprint {http://arxiv.org/abs/hep-th/0005232}
  {arXiv:hep-th/0005232} \BibitemShut {NoStop}%
\bibitem [{\citenamefont {Hall}(1994)}]{Hall1}%
  \BibitemOpen
  \bibfield  {author} {\bibinfo {author} {\bibfnamefont {Brian~C.}\
  \bibnamefont {Hall}},\ }\bibfield  {title} {\enquote {\bibinfo {title} {{The
  Segal-Bargmann Coherent State Transform for Compact Lie Groups}},}\
  }\href@noop {} {\bibfield  {journal} {\bibinfo  {journal} {J. Funct. Anal.},\
  }\textbf {\bibinfo {volume} {122}},\ \bibinfo {pages} {103--151} (\bibinfo
  {year} {1994})}\BibitemShut {NoStop}%
\bibitem [{\citenamefont {Hall}(2002)}]{Hall2}%
  \BibitemOpen
  \bibfield  {author} {\bibinfo {author} {\bibfnamefont {Brian~C.}\
  \bibnamefont {Hall}},\ }\bibfield  {title} {\enquote {\bibinfo {title}
  {{Geometric quantization and the generalized Segal-Bargmann transform for Lie
  groups of compact type}},}\ }\href@noop {} {\bibfield  {journal} {\bibinfo
  {journal} {Comm. Math. Phys.},\ }\textbf {\bibinfo {volume} {226}},\ \bibinfo
  {pages} {233--268} (\bibinfo {year} {2002})}\BibitemShut {NoStop}%
\bibitem [{\citenamefont {Ashtekar}\ \emph {et~al.}(1996)\citenamefont
  {Ashtekar}, \citenamefont {Lewandowski}, \citenamefont {Marolf},
  \citenamefont {Mourao},\ and\ \citenamefont {Thiemann}}]{Ashtekar:1994nx}%
  \BibitemOpen
  \bibfield  {author} {\bibinfo {author} {\bibfnamefont {Abhay}\ \bibnamefont
  {Ashtekar}}, \bibinfo {author} {\bibfnamefont {Jerzy}\ \bibnamefont
  {Lewandowski}}, \bibinfo {author} {\bibfnamefont {Donald}\ \bibnamefont
  {Marolf}}, \bibinfo {author} {\bibfnamefont {Jose}\ \bibnamefont {Mourao}}, \
  and\ \bibinfo {author} {\bibfnamefont {Thomas}\ \bibnamefont {Thiemann}},\
  }\bibfield  {title} {\enquote {\bibinfo {title} {{Coherent state transforms
  for spaces of connections}},}\ }\Doi {10.1006/jfan.1996.0018} {\bibfield
  {journal} {\bibinfo  {journal} {J. Funct. Anal.},\ }\textbf {\bibinfo
  {volume} {135}},\ \bibinfo {pages} {519--551} (\bibinfo {year} {1996})},\
  \Eprint {http://arxiv.org/abs/gr-qc/9412014} {arXiv:gr-qc/9412014}
  \BibitemShut {NoStop}%
\bibitem [{\citenamefont {Thiemann}(2001){\natexlab{c}}}]{Thiemann:2000bw}%
  \BibitemOpen
  \bibfield  {author} {\bibinfo {author} {\bibfnamefont {Thomas}\ \bibnamefont
  {Thiemann}},\ }\bibfield  {title} {\enquote {\bibinfo {title} {{Gauge field
  theory coherent states (GCS). I: General properties}},}\ }\Doi
  {10.1088/0264-9381/18/11/304} {\bibfield  {journal} {\bibinfo  {journal}
  {Class. Quant. Grav.},\ }\textbf {\bibinfo {volume} {18}},\ \bibinfo {pages}
  {2025--2064} (\bibinfo {year} {2001}{\natexlab{c}})},\ \Eprint
  {http://arxiv.org/abs/hep-th/0005233} {arXiv:hep-th/0005233} \BibitemShut
  {NoStop}%
\bibitem [{\citenamefont {Thiemann}\ and\ \citenamefont
  {Winkler}(2001){\natexlab{b}}}]{Thiemann:2000ca}%
  \BibitemOpen
  \bibfield  {author} {\bibinfo {author} {\bibfnamefont {T.}~\bibnamefont
  {Thiemann}}\ and\ \bibinfo {author} {\bibfnamefont {O.}~\bibnamefont
  {Winkler}},\ }\bibfield  {title} {\enquote {\bibinfo {title} {{Gauge field
  theory coherent states (GCS). II: Peakedness properties}},}\ }\Doi
  {10.1088/0264-9381/18/14/301} {\bibfield  {journal} {\bibinfo  {journal}
  {Class. Quant. Grav.},\ }\textbf {\bibinfo {volume} {18}},\ \bibinfo {pages}
  {2561--2636} (\bibinfo {year} {2001}{\natexlab{b}})},\ \Eprint
  {http://arxiv.org/abs/hep-th/0005237} {arXiv:hep-th/0005237} \BibitemShut
  {NoStop}%
\bibitem [{\citenamefont {Bahr}\ and\ \citenamefont
  {Thiemann}(2009)}]{Bahr:2007xn}%
  \BibitemOpen
  \bibfield  {author} {\bibinfo {author} {\bibfnamefont {Benjamin}\
  \bibnamefont {Bahr}}\ and\ \bibinfo {author} {\bibfnamefont {Thomas}\
  \bibnamefont {Thiemann}},\ }\bibfield  {title} {\enquote {\bibinfo {title}
  {{Gauge-invariant coherent states for Loop Quantum Gravity II: Non-abelian
  gauge groups}},}\ }\Doi {10.1088/0264-9381/26/4/045012} {\bibfield  {journal}
  {\bibinfo  {journal} {Class. Quant. Grav.},\ }\textbf {\bibinfo {volume}
  {26}},\ \bibinfo {pages} {045012} (\bibinfo {year} {2009})},\ \Eprint
  {http://arxiv.org/abs/0709.4636} {arXiv:0709.4636 [gr-qc]} \BibitemShut
  {NoStop}%
\bibitem [{\citenamefont {Bianchi}\ \emph
  {et~al.}(2010){\natexlab{c}}\citenamefont {Bianchi}, \citenamefont
  {Magliaro},\ and\ \citenamefont {Perini}}]{Bianchi:2010mw}%
  \BibitemOpen
  \bibfield  {author} {\bibinfo {author} {\bibfnamefont {Eugenio}\ \bibnamefont
  {Bianchi}}, \bibinfo {author} {\bibfnamefont {Elena}\ \bibnamefont
  {Magliaro}}, \ and\ \bibinfo {author} {\bibfnamefont {Claudio}\ \bibnamefont
  {Perini}},\ }\bibfield  {title} {\enquote {\bibinfo {title} {{Spinfoams in
  the holomorphic representation}},}\ }\href@noop {} { (\bibinfo {year}
  {2010}{\natexlab{c}})},\ \Eprint {http://arxiv.org/abs/1004.4550}
  {arXiv:1004.4550 [gr-qc]} \BibitemShut {NoStop}%
\bibitem [{\citenamefont {Wald}()}]{Wald:1984rg}%
  \BibitemOpen
  \bibfield  {author} {\bibinfo {author} {\bibfnamefont {Robert~M.}\
  \bibnamefont {Wald}},\ }\bibfield  {title} {\enquote {\bibinfo {title}
  {{General Relativity}},}\ }\href@noop {} {}\bibinfo {note} {Chicago, Usa:
  Univ. Pr. ( 1984) 491p}\BibitemShut {NoStop}%
\bibitem [{\citenamefont {Bojowald}\ and\ \citenamefont
  {Das}(2008){\natexlab{a}}}]{Bojowald:2007nu}%
  \BibitemOpen
  \bibfield  {author} {\bibinfo {author} {\bibfnamefont {Martin}\ \bibnamefont
  {Bojowald}}\ and\ \bibinfo {author} {\bibfnamefont {Rupam}\ \bibnamefont
  {Das}},\ }\bibfield  {title} {\enquote {\bibinfo {title} {{Canonical Gravity
  with Fermions}},}\ }\Doi {10.1103/PhysRevD.78.064009} {\bibfield  {journal}
  {\bibinfo  {journal} {Phys. Rev.},\ }\textbf {\bibinfo {volume} {D78}},\
  \bibinfo {pages} {064009} (\bibinfo {year} {2008}{\natexlab{a}})},\ \Eprint
  {http://arxiv.org/abs/0710.5722} {arXiv:0710.5722 [gr-qc]} \BibitemShut
  {NoStop}%
\bibitem [{\citenamefont {Bojowald}\ and\ \citenamefont
  {Das}(2008){\natexlab{b}}}]{Bojowald:2008nz}%
  \BibitemOpen
  \bibfield  {author} {\bibinfo {author} {\bibfnamefont {Martin}\ \bibnamefont
  {Bojowald}}\ and\ \bibinfo {author} {\bibfnamefont {Rupam}\ \bibnamefont
  {Das}},\ }\bibfield  {title} {\enquote {\bibinfo {title} {{Fermions in Loop
  Quantum Cosmology and the Role of Parity}},}\ }\Doi
  {10.1088/0264-9381/25/19/195006} {\bibfield  {journal} {\bibinfo  {journal}
  {Class. Quant. Grav.},\ }\textbf {\bibinfo {volume} {25}},\ \bibinfo {pages}
  {195006} (\bibinfo {year} {2008}{\natexlab{b}})},\ \Eprint
  {http://arxiv.org/abs/0806.2821} {arXiv:0806.2821 [gr-qc]} \BibitemShut
  {NoStop}%
\bibitem [{\citenamefont {Kodama}(1988)}]{Kodama:1988}%
  \BibitemOpen
  \bibfield  {author} {\bibinfo {author} {\bibfnamefont {Hideo}\ \bibnamefont
  {Kodama}},\ }\bibfield  {title} {\enquote {\bibinfo {title} {Specialization
  of ashtekar's formalism to bianchi cosmology},}\ }\Doi {10.1143/PTP.80.1024}
  {\bibfield  {journal} {\bibinfo  {journal} {Progress of Theoretical
  Physics},\ }\textbf {\bibinfo {volume} {80}},\ \bibinfo {pages} {1024--1040}
  (\bibinfo {year} {1988})}\BibitemShut {NoStop}%
\bibitem [{\citenamefont {Smolin}(2002)}]{Smolin:2002sz}%
  \BibitemOpen
  \bibfield  {author} {\bibinfo {author} {\bibfnamefont {Lee}\ \bibnamefont
  {Smolin}},\ }\bibfield  {title} {\enquote {\bibinfo {title} {{Quantum gravity
  with a positive cosmological constant}},}\ }\href@noop {} { (\bibinfo {year}
  {2002})},\ \Eprint {http://arxiv.org/abs/hep-th/0209079}
  {arXiv:hep-th/0209079} \BibitemShut {NoStop}%
\bibitem [{\citenamefont {Freidel}\ and\ \citenamefont
  {Smolin}(2004)}]{Freidel:2003pu}%
  \BibitemOpen
  \bibfield  {author} {\bibinfo {author} {\bibfnamefont {Laurent}\ \bibnamefont
  {Freidel}}\ and\ \bibinfo {author} {\bibfnamefont {Lee}\ \bibnamefont
  {Smolin}},\ }\bibfield  {title} {\enquote {\bibinfo {title} {{The
  linearization of the Kodama state}},}\ }\Doi {10.1088/0256-307X/21/8/002}
  {\bibfield  {journal} {\bibinfo  {journal} {Class. Quant. Grav.},\ }\textbf
  {\bibinfo {volume} {21}},\ \bibinfo {pages} {3831--3844} (\bibinfo {year}
  {2004})},\ \Eprint {http://arxiv.org/abs/hep-th/0310224}
  {arXiv:hep-th/0310224} \BibitemShut {NoStop}%
\end{thebibliography}%

\end{document}